# Sub-model aggregation for scalable eigenvector spatial filtering: Application to spatially varying coefficient modeling

Daisuke Murakami[1], Shonosuke Sugasawa[2], Hajime Seya[3], Daniel A. Griffith[4]

[1] Department of Statistical Data Science, Institute for Statistical Mathematics, Japan

[2] Graduate School of Economics, Keio University, Japan

[3] Departments of Civil Engineering, Graduate School of Engineering, Kobe University, Japan

[4] School of Economic, Political and Policy Sciences, The University of Texas at Dallas, USA

Abstract:

This study proposes a method for aggregating/synthesizing global and local sub-models for fast and flexible spatial regression modeling. Eigenvector spatial filtering (ESF) was used to model spatially varying coefficients and spatial dependence in the residuals by sub-model, while the generalized product-of-experts method was used to aggregate these sub-models. The major advantages of the proposed method are as follows: (i) it is highly scalable for large samples in terms of accuracy and computational efficiency; (ii) it is easily implemented by estimating sub-models independently first and aggregating/averaging them thereafter; and (iii) likelihood-based inference is available because the marginal likelihood is available in closed-form. The accuracy and computational efficiency of the proposed method are confirmed using Monte Carlo simulation experiments. This method was then applied to residential land price analysis in Japan. The results demonstrate the usefulness of this method for improving the interpretability of spatially varying coefficients. The proposed method is implemented in an R package spmoran (version 0.3.0 or later).

1. Introduction

In the era of big data, the statistical modeling of large spatial data has become increasingly important for regression analysis, prediction, uncertainty quantification, and other purposes (see Gelfand et al., 2010; Cressie and Wikle, 2015). The spatial regression model, whose computational cost rapidly grows as sample size increases, has been extended to enhance the computational efficiency as reviewed by Heaton et al. (2019). However, fast spatial regression model estimation typically assumes a relatively simple formulation with a linear mean function and distance-decaying covariance function to explain spatial dependence, which is one of the most basic properties of spatial data (Anselin, 2002). To make the spatial regression more flexible and scalable for large samples, this study considers extending eigenvector spatial filtering (ESF; Griffith, 2003; Griffith et al., 2019), which is a spatial regression approach.

ESF models spatial dependence using eigenvectors extracted from a spatial connectivity matrix. The eigenvectors corresponding to positive eigenvalues explain positive spatial dependence, whereas the reverse is true for negative spatial dependence (see Tiefelsdorf and Griffith, 2007). Using this property, ESF models positive spatial dependence, which is dominant in most real-world cases, using a linear combination of $L$ eigenvectors paired with their positive eigenvalues. ESF has been used to analyze urbanization (e.g., Yan et al., 2021; Peng and Inoue, 2022; Murakami and Seya, 2022), demography (e.g., Thayn and Simanis, 2013), environment (e.g., Mainali and Chang, 2021; Park et al., 2022), ecology (e.g., Diniz-Filho and Bini, 2005; Griffith and Peres-Neto, 2006) and epidemiology (e.g., Griffith and Li, 2021) data.

Still, for large samples, ESF has difficulties balancing (i) computational cost and (ii) modeling accuracy, as demonstrated in Section 2.2. Regarding (i), ESF requires an eigen-decomposition of an $N \times N$ connectivity matrix whose computational complexity rapidly grows in cubic order with respect to the sample size $N$. Moreover, regression using $L$ eigenvectors can be slow if $L$ is large, especially if it involves a stepwise selection procedure.

Regarding (ii) accuracy, an ESF model is categorized as a low-rank model that describes a spatial process using $L$ ($< N$) basis functions. Stein (2014) showed a degeneracy problem in that low rank modeling can cause an over-smoothing of a spatial process outcome if $L << N$; ESF also can perform poorly for very large samples, as other low rank models do, including fixed rank kriging (Cressie and Johannesson, 2008), predictive process models (Banerjee et al., 2008), and generalized additive models (Wood, 2017), all of which have been used for massively large samples.

ESF has been extended to large samples through global or local approximations. Murakami and Griffith (2019a, b) propose a global method that approximates a limited number of eigenvectors to explain large-scale map patterns, and show its applicability for modeling spatially varying coefficients (SVCs) using millions of samples while assuming $L = 200$. However, their approach encounters the aforementioned difficulty in terms of (i) accuracy because of the small $L$ that induces the degeneracy/over-smoothing problem in SVCs.[1] In contrast, the local approximations devised by Griffith and Chun (2019) and Yang et al. (2020) divide the study area into sub-regions and apply ESF for each region. These local approaches mitigate the degeneracy problem as well as reduce computation costs. Yet, because they consider only local sub-samples, their estimation can be unstable when estimating complex spatial models, such as SVC specifications. In addition, a spatial process can be discontinuous at the border separating subregions. Thus, local approximation has room to enhance (i) modeling accuracy.

The objective of this study is to make the ESF fast, accurate, and flexible for large-scale spatial modeling. To achieve this, we employed sub-model aggregation (Hinton, 2002; Rullière et al., 2018), which is an ensemble learning technique. Using this technique, we aggregated/combined global and local ESF models, which were estimated a priori, to build a complex but computationally efficient model. Among the ESF model specifications, we focus on a random-effects ESF because of its better stability (see Murakami and Griffith, 2015).

The remining sections are organized as follows: Sections 2 and 3 respectively introduce the random-effects ESF and sub-model aggregation. Section 4 describes the development of our proposed method. Section 5 compares our method with conventional spatial models using Monte Carlo simulation experiments, and Section 6 applies the method to a residential land price analyses in Japan. Finally, Section 7 concludes the study.

---

[1] The $L$ eigenvectors may include ones that are not strongly correlated with the response variable, introducing noise into the SVCs and reducing accuracy of ESF predictions. The random-effects ESF (Murakami and Griffith, 2015), which we use later, mitigates this problem through estimation that imposes a stronger regularization on the coefficients of the eigenvectors corresponding to smaller eigenvalues, which explain less variation.

2. Random-effects ESF and spatial regression

This section introduces a random-effects ESF-based SVC model (Murakami et al., 2017) that considers spatial dependence in the residuals and regression coefficients. Section 2.1 introduces the model, and Section 2.2 explains its scalability for large samples.

2.1. The model

Suppose that $y(s_1), ..., y(s_N)$ are the observations of the response/dependent variables at $N$ sample sites $s_1, ..., s_N$ distributed over a study region $D \subset \mathcal{R}^2$. $y(s_i)$ is specified as follows:

$$y(s_i) = \beta_0(s_i) + \sum_{k=1}^{K} x_k(s_i)\beta_k(s_i) + \varepsilon(s_i), \quad \varepsilon(s_i) \sim N(0, \sigma^2), \quad (1)$$

where $x_k(s_i)$ is the observation of the $k$-th explanatory variable and $\varepsilon(s_i)$ is the random noise with variance $\sigma^2$. The $k$-th SVC $\beta_k(s_i)$ is defined using a random effects ESF model, as follows (see Murakami et al., 2017):

$$\beta_k(s_i) = b_k + \sum_{l=1}^{L} e_l(s_i)\, v_l(\boldsymbol{\theta}_k) u_{k,l}, \qquad u_{k,l} \sim N(0, \sigma^2), \tag{2}$$

representing [mean: $b_k$] + [zero mean spatial process: $\sum_{l=1}^{L} e_l(s_i)\, v_l(\boldsymbol{\theta}_k) u_{k,l}$], where parameters $\boldsymbol{\theta}_k = [\alpha_k, \tau_k^2, \sigma^2]'$. $\mathbf{e}_l = [e_l(s_1), \ldots, e_l(s_N)]'$ and $\lambda_l$ are the $l$-th eigenvector and eigenvalue of a doubly centered connectivity matrix **MCM** with $\mathbf{M} = \mathbf{I} - \mathbf{1}\mathbf{1}'/N$ being a centering matrix in which **I** is an identity matrix and **1** is a vector of ones. **C** is a $N \times N$ spatial proximity matrix across sample sites with zero diagonal cell entries. In this paper's study, the $(i, j)$-th element is given by $c(d_{ij}) = \exp(-d_{ij}/r)$, where $d_{ij}$ is the Euclidean distance between sample sites and the range parameter $r$ is given by the longest distance in the minimum spanning tree connecting the sample sites, following Dray et al. (2006) and Murakami and Griffith (2015). $L$ eigenvectors satisfying $\lambda_l > 0$ explain positive spatial dependence in terms of the Moran coefficient (Moran, 1950), whereas those corresponding to negative eigenvalues explain negative dependence. We considered the former for modeling smoothly varying coefficient processes, which have been assumed in SVC modeling studies (e.g., Fotheringham et al., 2002).

The variance function $v_l(\boldsymbol{\theta}_k) = \frac{\tau_k}{\sigma} \lambda_l^{\alpha_k/2}$ characterizes the map pattern of $\beta_k(s_i)$. A larger $\tau_k^2$ means greater variation while $\tau_k^2 = 0$ means no variation in $\beta_k(s_i)$, the latter rendering a constant coefficient $\beta_k(s_i) = b_k$. A large-scale parameter $\alpha_k$ yields a large-scale map pattern. As $\alpha_k \to \infty$, the Moran coefficient of $\beta_k(s_i)$ approaches its maximum possible value indicating the strongest positive dependence available (see Griffith, 2003).

As $\alpha_k \to 0$, the Moran coefficient approaches zero, implying no spatial dependence. Thus, the $\alpha_k$ parameter controls the scale and Moran coefficient (Murakami and Griffith, 2020).

The SVC model is estimated by the following steps: (i) variance parameters $\{\tau_1^2, \dots, \tau_K^2, \alpha_1, \dots, \alpha_K\}$ are estimated by maximizing the marginal likelihood (see Section 4); and then (ii) given these parameters, the SVCs are predicted (see Murakami et al., 2017 for further detail).

2.2. Scalability of ESF

The ESF requires eigen-decomposition and parameter estimation, both of which can be slow for large samples. However, the computational cost can be reduced by approximating the eigenvectors and limiting the number of constructed eigenvectors, such that $L$ = 200 following Murakami and Griffith (2019a). Unfortunately, a small $L$ can deteriorate modeling accuracy if the true SVCs have small-scale patterns (Stein, 2014). This misspecification might be the reason why the ESF-based SVC model is less accurate in some scenarios, as reported by Dambon et al. (2021) and Fan and Huang (2022).

To illustrate this problem, let us fit an ESF-based SVC model [Eqs. (1)-(2)] with synthetic data generated on a 80 × 80 regularly spaced grid of points, and compare the fitted SVCs with geographically weighted regression (GWR; Fotheringham et al., 2002), which is a popular SVC modeling method. The simulated data are generated using a Gaussian linear regression model with a spatially varying intercept $\beta_0(s_i)$ and SVCs $\beta_1(s_i)$ and $\beta_2(s_i)$ at $i$-grid $s_i$; the same model generated synthetic data for Section 5.2. Small-scale map patterns are assumed for the SVCs (Figure 1 (a)).

Figure 1 maps $\beta_1(s_i)$ estimated from GWR and ESF with $L$ = 200. Relative to GWR, which does not suffer from a degeneracy problem, the coefficients estimated from the ESF tend to display an excessively smooth map pattern due to the small $L$. Table 1 compares the root mean squared error (RMSE) for the SVCs as well as computation time. As $L$ decreases, ESF becomes faster, but less accurate because the over-smoothing problem gets more severe. Although estimation accuracy improves as $L$ increases, the necessary computation time rapidly grows.

In short, the ESF-based SVC model struggles to balance computational efficiency and modeling accuracy for large samples. To overcome this limitation, the next section considers synthesizing multiple ESF models using a technique called sub-model aggregation.

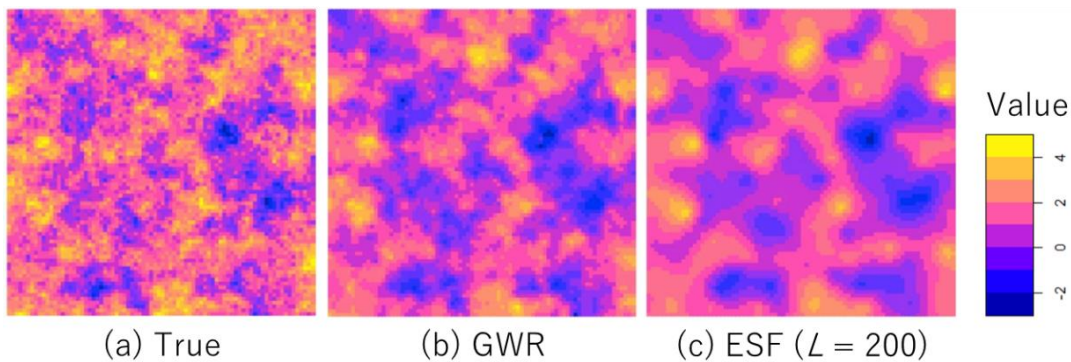

Figure 1: True and estimated/predicted SVCs ($\beta_1(s_i)$) in a preliminary analysis.

Table 1: Preliminary comparison of accuracy and computation time in SVC modeling.[1)]

|  | L | RMSE $\beta_0(s_i)$ | $\beta_1(s_i)$ | $\beta_2(s_i)$ | Computation time (seconds) |
|---|---|---|---|---|---|
| ESF | 50 | 0.84 | 1.67 | 0.81 | 1.31 |
|  | 200 | 0.69 | 1.36 | 0.71 | 18.52 |
|  | 500 | 0.63 | 1.23 | 0.67 | 244.26 |
|  | 1000 | 0.60 | 1.15 | 0.66 | 1254.50 |
| GWR |  | 0.65 | 1.15 | 0.78 | 83.82 |

[1)] Simulation data is generated on a 80 × 80 grids using a Gaussian regression model with spatially varying intercept $\beta_0(s_i)$ and two SVCs, $\beta_1(s_i)$ and $\beta_2(s_i)$. The standard deviations of $\beta_1(s_i)$ and $\beta_2(s_i)$ are 2 and 1 respectively, assuming that $\beta_1(s_i)$ is more influential than $\beta_2(s_i)$. See Section 5.2 for further details about these data generating process.

## 3. Sub-model aggregation

This section briefly introduces sub-model aggregation, which is used to develop our model in Section 4.

### 3.1. Basics

Sub-model aggregation has been used to approximate Gaussian processes (Rasmussen, 2004). Popular aggregation methods include the product-of-experts (PoE; Hinton, 2002), generalized PoE (Cao and Fleet, 2014), Bayesian committee machine (BCM; Tresp, 2000), and robust BCM (Deisenroth and Ng, 2015).

Let us partition the samples $y(s_1), \dots, y(s_N)$ into $C$ sub-samples via spatial clustering, and assume a sub-model $M_c$ for the $c$-th sub-sample. Sub-model aggregation approximates the predictive distribution of $y(s_i)$ given all sub-models $M_{1:C} = [M_1, \dots, M_C]'$ as follows:

$$p(y(s_i)|M_{1:C}) = \frac{1}{Z} \prod_{c=1}^{C} p(y(s_i)|M_c)^{w_c(s_i)}, \qquad (3)$$

where the (robust) BCM explicitly models the normalizing term $Z = p(y(s_i)|M_{1:C})^{\Sigma_c w_c(s_i) - 1}$ whereas the (generalized) PoE does not; $p(y(s_i)|M_c)$ is the predictive distribution of $y(s_i)$ given the $c$-th sub-model.

Eq. (3) synthesizes the sub-models by considering a known prior weight $w_c(s_i)$.

Sub-model $M_c$ with a large weight $w_c(s_i)$ has a strong influence at site $s_i$, whereas any sub-model with $w_c(s_i) = 0$ has no influence at the site. PoE and BCM assume $w_c(s_i) = 1$ and hence ignore the weight, whereas the generalized PoE and robust BCM specify $w_c(s_i)$ based upon some criteria. Generalized PoE is identical to a robust BCM if the weights are normalized such that $\sum_c w_c(s_i) = 1$. This is because the normalizing term becomes $Z = 1$, and thus Eq. (3) reduces to

$$p(y(s_i)|M_{1:C}) = \prod_{c=1}^{C} p(y(s_i)|M_c)^{w_c(s_i)}. \qquad (4)$$

3.2. Generalized product-of-experts

PoE, BCM, and robust BCM assume common parameters and conditional independence across sub-models, meaning that subsamples cannot overlap. Exceptionally, the generalized PoE, which is strongly motivated by a log-opinion pooling framework, allows for different parameters across sub-models, and hence the sub-samples can overlap (Cao, 2018; Tautvaišas and Žilinskas, 2022). Therefore, the generalized PoE is flexible.

The generalized PoE has been used to aggregate Gaussian models because Eq. (4) is a closed-form expression in that case (e.g., Deisenroth and Ng, 2015). Specifically, the $c$-th sub-model $M_c$ is defined as follows:

$$y(s_i)|M_c \sim N\big(\mu_c(s_i), \sigma_c^2(s_i)\big), \quad (5)$$

with mean $\mu_c(s_i)$ and variance $\sigma_c^2(s_i)$, the aggregated model $M_{1:c}$ yields

$$y(s_i)|M_{1:c} \sim N\big(\mu_*(s_i), \sigma_*^2(s_i)\big), \quad (6)$$

where

$$\mu(s_i) = \frac{\sum_c w_c(s_i; \sigma_c^2)\mu_c(s_i)}{\sum_c w_c(s_i; \sigma_c^2)}, \quad (7)$$

$$\sigma^2(s_i) = \frac{1}{\sum_c w_c(s_i; \sigma_c^2)}. \quad (8)$$

The mean $\mu(s_i)$ equals the average of $\mu_1(s_i), \dots, \mu_C(s_i)$ weighted by $w_c(s_i; \sigma_c^2) = \frac{w_c(s_i)}{\sigma_c^2(s_i)}$. The posterior weight $w_c(s_i; \sigma_c^2)$ is given by the upward adjustment of the prior weight $w_c(s_i)$ for sub-models with a small predictive error variance $\sigma_c^2(s_i)$ at site $s_i$. The aggregated model can be easily obtained by estimating the $C$ sub-models given by Eq. (5) individually first, and substituting the predictive mean $\mu_c(s_i)$ and error variance $\sigma_c^2(s_i)$ into Eqs. (6)-(8) after that (see Figure 2).

Despite its simplicity and practicality of this procedure, the application of the generalized PoE remains limited in spatial statistics. To the best of the authors' knowledge, the generalized PoE has never been extended to either SVC or ESF modeling.

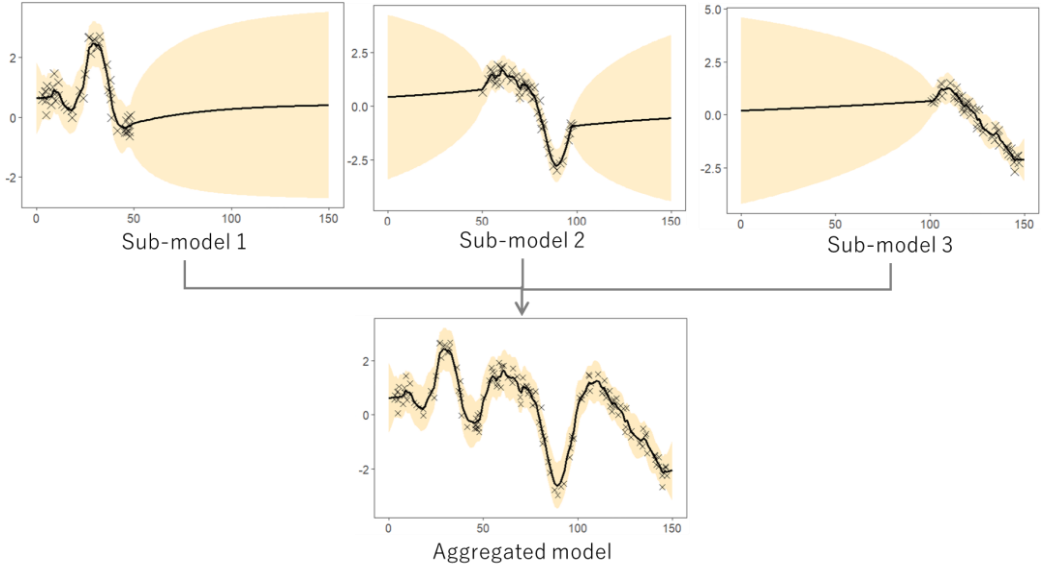

Figure 2: Generalized PoE. (×: sample; solid line: predictive mean; yellow area: 95 % confidence interval). Synthetic samples are generated across one-dimensional space [0, 150] using a Gaussian process with mean zero, variance $c_{ii} = 1.2^2$, and covariance decay with respect to $c_{ij} = \exp(-\frac{d_{ij}}{50})$. The samples are divided into three sub-samples, as illustrated. Sub-model Eq. (5) were fitted by the subsamples (top), and then aggregated using Eqs. (6)-(9) (bottom) to model the predicted mean and variance of the entire study area.

4. Proposed method

This section extends the ESF-based SVC models introduced in Section 2 to enhance the flexibility and scalability. After explaining the assumptions in Section 4.1,

Section 4.2 introduces the sub-models, and Section 4.3 explains how to aggregate them. Section 4.4 explains how to estimate the corresponding aggregated model.

4.1. Assumptions: prior weights and clusters

A generalized PoE was used to aggregate the sub-models. We divided the study area into $C - 1$ clusters and assumed sub-models for each cluster. While many algorithms or rules are available to determine the clusters, we used the *k*-means algorithm because of its computational efficiency.[2]

The prior weight $w_c(s_i)$ was specified in two ways: First, it was specified as

$$w_c(s_i) = \begin{cases} 1 & if\ s_i \in c \\ 0 & otherwise \end{cases}, \tag{9}$$

where $s_i \in c$ indicates that the site $s_i$ is in the *c*-th cluster. Given Eq. (9), the *c*-th sub-model only influences the *c*-th cluster. The cluster-wise ESF of Griffith and Chun (2019) and Yang et al. (2020) implicitly assume this. Unfortunately, this prior weight assumes independence across clusters, which makes a map pattern of the SVCs discontinuous near cluster boundaries.

To mitigate this problem, we considered another weight, assuming spatial dependence from the $c$-th sub-model to the neighbors, decaying with according to the distance $d_{i(c)}$ from the nearest sample site in the $c$-th cluster:

$$w_c(s_i) = \frac{w_{0,c}(s_i)}{\sum_c w_{0,c}(s_i)}, \quad w_{0,c}(s_i) = \begin{cases} c(d_{i(c)}) & if\ d_{i(c)} < r_c^{max} \\ 0 & otherwise \end{cases}, \quad (10)$$

where $c(d_{i(c)}) = \exp(-d_{i(c)}/r_c)$ is the same function used to model the spatial dependence in the $c$-th sub-model, with $r_c$ being a range parameter (see Section 2.1). $r_c^{max}$ is the threshold distance imposed for rapid computation. We assumed $r_c^{max} = 2.2r_c$. Because 90 % of the weights disappear at that distance,[3] thresholding only has a marginal influence on the modeling result. With Eq. (10), the $c$-th sub-model influences not only in the $c$-th cluster, but also its neighbors. The resulting SVC process becames smooth after the sub-model aggregation.

---

[2] As suggested by Masoudnia and Ebrahimpour (2012), such an approach determining clusters before model estimation allows each sub-model to effectively learn local properties in the corresponding cluster, leading to a better generalization ability than approaches optimizing clusters and a regression model simultaneously. Therefore, we rely on the a priori clustering approach.

Still, the $C - 1$ local sub-models may overlook the global patterns across clusters. To address this problem, the $C$-th sub-model was defined as a global model that considers all samples, assuming $w_{0,C}(s_i) = 1$. The next subsection explains the local and global sub-models.

---

[3] As explained in Section 3.2, mean $\mu(s_i)$ of the aggregated model is given by the average of $\mu_c(s_i)$ weighted by $\frac{w_1(s_i;\sigma_1^2)}{w_1(s_i;\sigma_1^2)+w_2(s_i;\sigma_2^2)}$. Consider a border separating clusters 1 and 2 where the other sub-models are far from the border area. In a homoscedastic case with $\sigma_1^2(s_i) = \sigma_2^2(s_i)$, the relative weight of the first model $M_1$ at a site $s_i$ in the second cluster $(\exp\left(-\frac{d_{i(2)}}{r_2}\right) = \exp(-\frac{0}{r_2}) = 1)$ becomes $\frac{w_1(s_i;\sigma_1^2)}{w_1(s_i;\sigma_1^2)+w_2(s_i;\sigma_2^2)} = \frac{1}{\frac{\exp(-d_{i(1)}/r_1)}{\sigma_1^2(s_i)}+\frac{1}{\sigma_2^2(s_i)}} \frac{\exp(-d_{i(1)}/r_1)}{\sigma_1^2(s_i)} = \frac{\exp(-d_{i(1)}/r_1)}{\exp(-d_{i(1)}/r_1)+1} = \frac{1}{1+\exp(d_{i(1)}/r_1)}$, which is a logistic function. At the threshold $d_{i(1)} = 2.2r_1$, the relative weight for the first model becomes $\frac{1}{1+\exp(2.2)} \approx 0.100$.

### 4.2. Sub-models

In each cluster, the local sub-models $M_1, \ldots, M_{C-1}$ were assumed to be

$$y(s_i)|M_c \sim N\left(\sum_{k=0}^{K} x_k(s_i)\beta_{k,c}(s_i), \sigma_c^2\right), \tag{11}$$

where $\sigma_c^2$ denotes the noise variance. Following Eq. (2), SVC was defined as

$$\beta_{k,c}(s_i) = b_{k,c} + \sum_{l=0}^{L_c} e_l(s_i)v_l(\boldsymbol{\theta}_{k,c})u_{l,c}, \qquad u_{l,c} \sim N(0, \sigma_c^2), \tag{12}$$

where $L_c$ denotes the number of eigenvectors and $v_l(\boldsymbol{\theta}_{k,c}) = \frac{\tau_{k,c}}{\sigma_c}\lambda_l^{\alpha_{k,c}/2}$ determines the variance and spatial scale of the SVC using parameters $\boldsymbol{\theta}_{k,c} = [\alpha_{k,c}, \tau_{k,c}^2, \sigma_c^2]'$. To derive an ensuing likelihood, the variance of $u_{l,c}$ was assumed to be the same as that of $y(s_i)$. Because the variance is rescaled by multiplying $v_l(\boldsymbol{\theta}_{k,c})$ in Eq. (12), this assumption does not affect the resulting SVC estimates. $\beta_{k,c}(s_i)$ becomes a constant if $\tau_{k,c} = 0$. The *c*-th local sub-model [Eqs. (11)-(12)] is given for sub-samples satisfying $w_c(s_i) > 0$. All the $L_c$ eigenvectors corresponding to the positive eigenvalues are considered to explain the positive spatial dependence in the *c*-th cluster.

The global sub-model $M_C$ is also defined by Eqs. (11)-(12), but estimated using all the samples because $w_{0,C}(s_i) = 1$ for all *i*. Therefore, the eigenvectors corresponding to the 200 largest positive eigenvalues were considered for modeling large-scale positive dependency map patterns based on the suggestion that the most positively correlated variations explained by the Moran coefficient are described by 200 eigenvectors, even for large samples such as $N$ = 100,000 (see Appendix of Murakami and Griffith, 2019a).

In distribution form, the $c$-th sub-model in Eqs. (11)-(12) at site $s_i$ can be expressed as

$$p(y(s_i), \mathbf{u}_c | \mathbf{b}_c, \boldsymbol{\theta}_c, \sigma_c^2) \propto p(y(s_i) | \mathbf{u}_c, \mathbf{b}_c, \boldsymbol{\theta}_c, \sigma_c^2) p(\mathbf{u}_c | \sigma_c^2), \quad (13)$$

where $\mathbf{b}_c = [b_{1,c}, \dots, b_{K,c}]'$, $\mathbf{u}_c = [u_{1,c}, \dots, u_{L,c}]'$, and $\boldsymbol{\theta}_c = [\boldsymbol{\theta}_{c,1}, \dots, \boldsymbol{\theta}_{c,K}]'$. The sub-models introduced in this section are used to construct the aggregated model in the next section.

4.3. Aggregated model

This section explains our model, which aggregates the sub-models introduced in Section 4.2. Sections 4.3.1 and 4.3.2 respectively present the aggregated model in a distribution and regression form.

4.3.1. A distribution form

Section 4.2 introduces $C$ sub-models $M_1, \dots, M_C$ each of which has the distribution $p(y(s_i), \mathbf{u}_c | \mathbf{u}_c, \mathbf{b}_c, \boldsymbol{\theta}_c, \sigma_c^2)$. This section combines these sub-models to

construct an aggregated model $p(\mathbf{y}, \mathbf{u}_{1:C}|\mathbf{b}_{1:C}, \boldsymbol{\theta}_{1:C}, \sigma^2_{1:C})$, where $\mathbf{y} = [y(s_1), ..., y(s_N)]'$, $\mathbf{u}_{1:C} = [\mathbf{u}_1, ..., \mathbf{u}_C]'$, and $\mathbf{b}_{1:C}, \boldsymbol{\theta}_{1:C}, \sigma^2_{1:C}$ are similarly defined. Based on Eq. (13), we first aggregate $p(y(s_i)|\mathbf{u}_c, \mathbf{b}_c, \boldsymbol{\theta}_c, \sigma^2_c)$ and $p(\mathbf{u}_c|\sigma^2_c)$ respectively, and then combine them after that.

Regarding $p(y(s_i)|\mathbf{u}_c, \mathbf{b}_c, \boldsymbol{\theta}_c, \sigma^2_c)$, we consider the following generalized PoE-type aggregation:

$$p(y(s_i)|\mathbf{u}_{1:C}, \mathbf{b}_{1:C}, \boldsymbol{\theta}_{1:C}, \sigma^2_{1:C}) = \prod_{c=1}^{C} p(y(s_i)|\mathbf{u}_c, \mathbf{b}_c, \boldsymbol{\theta}_c, \sigma^2_c)^{w_c(s_i)}, \quad (14)$$

where $\sum_c w_c(s_i) = 1$. Because the sub-models are Gaussians, the aggregate model is also Gaussian, just like the usual generalized PoE. The distribution of $\mathbf{y} = [y(s_1), ..., y(s_N)]'$ is expressed as:

$$\begin{aligned} p(\mathbf{y}|\mathbf{u}_{1:C}, \mathbf{b}_{1:C}, \boldsymbol{\theta}_{1:C}, \sigma^2_{1:C}) &\propto \prod_{i=1}^{N} \prod_{c=1}^{C} p(y(s_i)|\mathbf{u}_c, \mathbf{b}_c, \boldsymbol{\theta}_c, \sigma^2_c)^{w_c(s_i)}, \\ &= \prod_{c=1}^{C} \prod_{i_c=1}^{N_c} p(y(s_{i_c})|\mathbf{u}_c, \mathbf{b}_c, \boldsymbol{\theta}_c, \sigma^2_c)^{w_c(s_{i_c})}, \end{aligned} \quad (15)$$

where $i_c \in \{1, ..., N_c\}$ denotes the samples satisfying $w_c(s_i) > 0$. The second line is obtained by using $p(y(s_{i_c})|\mathbf{u}_c, \mathbf{b}_c, \boldsymbol{\theta}_c, \sigma^2_c)^0 = 1$.

Regarding the random coefficients vector $\mathbf{u}_c$, we assume the following generalized PoE-based aggregation:

$$p(\mathbf{u}_{1:C}|\sigma_{1:C}^2) \propto \prod_{c=1}^{C} p(\mathbf{u}_c|\sigma_c^2)^{\frac{W_c}{N_c}}, \tag{16}$$

where $W_c = \sum_{i_c=1}^{N_c} w_c(s_{i_c})$ is the total weight of the $c$-th sub-model, which takes a value between zero and $N_c$. The multiplier $W_c/N_c$ downweighs the $c$-th sub-model $p(\mathbf{u}_c|\sigma_c^2)$ according to $W_c$. If $W_c = 0$, $p(\mathbf{u}_c|\sigma_c^2)$ becomes flat and has no impact. In contrast, if $W_c = N_c$, which occurs if $w_c(s_{i_c}) = 1$ for all samples in the $c$-th cluster, $p(\mathbf{u}_c|\sigma_c^2)$ is the only sub-model influencing the $c$-th subsample.

By multiplying Eqs. (15) and (16), the aggregated model can be expressed as follows:

$$\begin{aligned} p(\mathbf{y}, \mathbf{u}_{1:C}|\mathbf{b}_{1:C}, \boldsymbol{\theta}_{1:C}, \sigma_{1:C}^2) &\propto p(\mathbf{y}|\mathbf{u}_{1:C}, \mathbf{b}_{1:C}, \boldsymbol{\theta}_{1:C}, \sigma_{1:C}^2) p(\mathbf{u}_{1:C}|\sigma_{1:C}^2) \\ &= \prod_{c=1}^{C} \prod_{i_c=1}^{N_c} p(y(s_{i_c})|\mathbf{u}_c, \mathbf{b}_c, \boldsymbol{\theta}_c, \sigma_c^2)^{w_c(s_{i_c})} p(\mathbf{u}_c|\sigma_c^2)^{W_c} \end{aligned} \tag{17}$$

### 4.3.2. A regression form

As detailed in Appendix 1, the aggregated SVC model in Eq. (17) can be expressed as follows:

$$y(s_i)|M_{1:C} \sim N\left(\sum_{k=1}^{K} x_k(s_i)\beta_{k*}(s_i), \sigma_*^2(s_i)\right), \tag{18}$$

where $\sigma_*^2(s_i) = \left(\sum_c \frac{w_c(s_i)}{\sigma_c^2(s_i)}\right)^{-1}$. The aggregated SVC $\beta_{k*}(s_i)$ is given by

$$\beta_{k*}(s_i) = \frac{1}{\sum_{c=1}^{C} w_c(s_i; \sigma_c^2)} \sum_{c=1}^{C} w_c(s_i; \sigma_c^2) \beta_{k,c*}(s_i). \tag{19}$$

Eq. (19) averages the coefficients $\beta_{k,1*}(s_i), \ldots, \beta_{k,C*}(s_i)$ of each sub-model by using the weight $w_c(s_i; \sigma_c^2) = \frac{w_c(s_i)}{\sigma_c^2(s_i)}$, where

$$\beta_{k,c*}(s_i) = b_{k,c} + \sum_{l=0}^{L_c} e_l(s_i) v_l(\boldsymbol{\theta}_{k,c}) u_{l,c}, \qquad u_l \sim N\left(0, \frac{N_c}{W_c} \sigma_c^2\right). \tag{20}$$

Here, the variance decreases as the model weight $W_c$ increases. Figure 3 illustrates the procedure for aggregating the sub-models SVC $\beta_{k,1*}(s_i), \ldots, \beta_{k,C*}(s_i)$ to construct aggregated SVC $\beta_{k*}(s_i)$.

Note that the sub-model SVC is replaced with a sub-model-wise constant by assuming $\beta_{k,c*}(s_i) = b_{k,c}$ instead of Eq. (20). This property is useful for reducing model complexity and stabilizing estimation. Furthermore, a global constant is available by substituting $\beta_{k,c*}(s_i) = b_{k,c}$ into Eq. (19), and averaging them across samples. In other words, when assuming a constant coefficient for the *k*-th explanatory variable, this value becomes

$$b_{k*} = \frac{1}{\sum_{i=1}^{N} \sum_{c=1}^{C} w_c(s_i; \sigma_c^2)} \sum_{i=1}^{N} \sum_{c=1}^{C} w_c(s_i; \sigma_c^2) b_{k,c}. \tag{21}$$

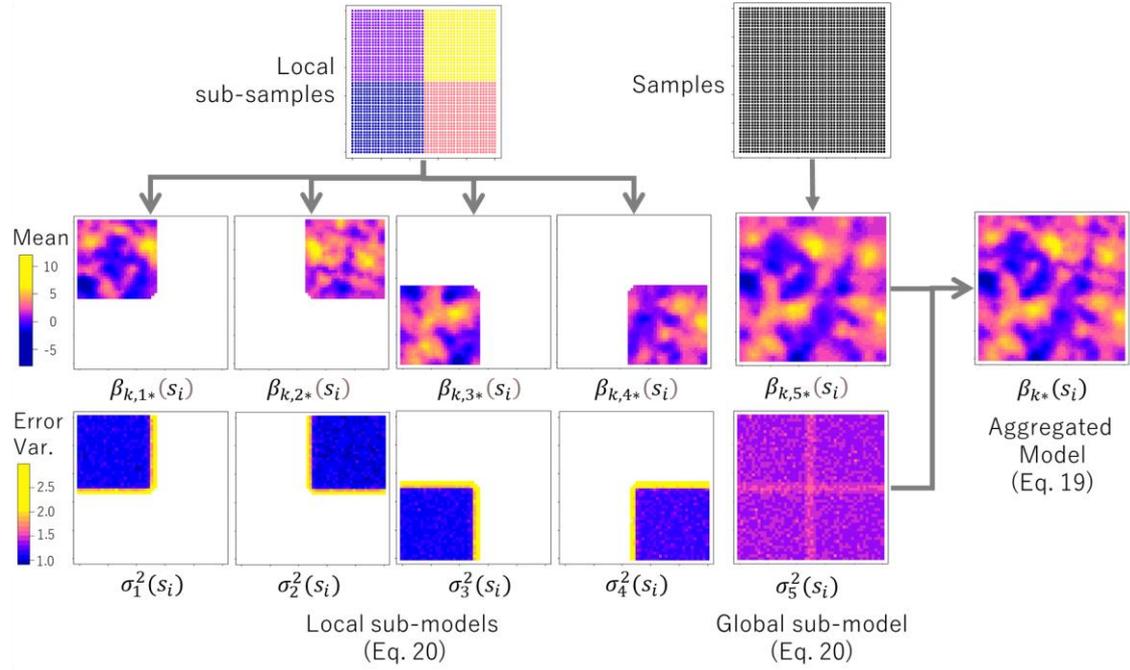

Figure 3: An image of aggregated SVC assuming four local subsamples (top left). Four local sub-models and one global sub-model are assumed to have their own SVCs $\beta_{k,1*}(s_i), \ldots, \beta_{k,5*}(s_i)$. The SVCs are averaged using Eq. (19) to obtain the aggregated SVC $\beta_{k*}(s_i)$. Therein, the SVCs are averaged site by site by weighting each sub-model with the posterior weight $w_c(s_i; \sigma_c^2) = \frac{w_c(s_i)}{\sigma_c^2(s_i)}$ that increases for sub-models with a large prior weight $w_c(s_i)$ and a small error variance $\sigma_c^2(s_i)$.

4.4. Estimation

Two types of marginal likelihood are available for estimating the aggregated model in Eqs. (18)-(20). The first does not marginalize the fixed coefficient vector $\mathbf{b}_{1:C}$ whereas the second marginalizes them (see Appendices 2 and 3). We use the latter, which is also called a restricted likelihood, which frequently is preferred for accurately estimating variance parameters (Bates, 2010). As detailed in Appendices 2 and 3, the marginal likelihood becomes

$$l(\boldsymbol{\theta}_{1:C}) = \sum_{c=1}^{C} l(\boldsymbol{\theta}_c),$$

$$l(\boldsymbol{\theta}_c) = -\frac{W_c - K}{2} \log\left(2\pi \frac{\hat{\boldsymbol{\varepsilon}}'_c \ddot{\mathbf{W}}_c \hat{\boldsymbol{\varepsilon}}_c + \hat{\mathbf{u}}'_c \hat{\mathbf{u}}_c}{N_c - K}\right) - \frac{1}{2} \log|\mathbf{M}_c| - \frac{W_c}{2}.$$

(22)

$\ddot{\mathbf{W}}_c = \frac{N_c}{W_c} \mathbf{W}_c$ , $\hat{\boldsymbol{\varepsilon}}_c = \mathbf{y}_c - \mathbf{X}_c \hat{\mathbf{b}}_c - \mathbf{E}_c \mathbf{V}_c(\boldsymbol{\theta}_c) \hat{\mathbf{u}}_c$ , and $\mathbf{M}_c = \begin{bmatrix} \mathbf{X}'_c \ddot{\mathbf{W}}_c \mathbf{X}_c & \mathbf{X}'_c \ddot{\mathbf{W}}_c \mathbf{E}_c \mathbf{V}_c(\boldsymbol{\theta}_c) \\ \mathbf{V}_c(\boldsymbol{\theta}_c) \mathbf{E}'_c \ddot{\mathbf{W}}_c \mathbf{X}_c & \mathbf{V}_c(\boldsymbol{\theta}_c) \mathbf{E}'_c \ddot{\mathbf{W}}_c \mathbf{E}_c \mathbf{V}_c(\boldsymbol{\theta}_c) + \mathbf{I}_c \end{bmatrix}$, where $\mathbf{y}_c$ is an $N_c \times 1$ vector of observed response variables in the $c$-th cluster, $\mathbf{X}_c$ is an $N_c \times K$ matrix of observed explanatory variables, $\mathbf{E}_c$ is an $N_c \times KL_c$ matrix whose $(i, (k-1)L + l)$-th element equals $x_k(s_i)e_l(s_i)$, $\mathbf{V}_c(\boldsymbol{\theta}_c)$ is a diagonal matrix whose $(i, (k-1)L + l)$-th element equals $v_l(\boldsymbol{\theta}_{k,c})$, and $\mathbf{I}_c$ is an identity matrix; vectors $\hat{\mathbf{b}}_c$ and $\hat{\mathbf{u}}_c$ are the estimators/predictors of $\mathbf{b}_c$ and $\mathbf{u}_c$ that maximize the likelihood function.

Note that the conventional random effects ESF-based SVC model [Eqs. (1)-(2)] assume $C = 1$, $w_c(s_{i_c}) = 1$, and $N_c = N$, meaning that $W_c = \sum_{i_c=1}^{N_c} w_c(s_{i_c}) = N$ and $\ddot{\mathbf{W}}_c = \mathbf{I}$. By substituting these quantities, Eq. (22), reduces the marginal likelihood of a basic random effects ESF. The proposed model Eqs. (18)-(20) includes the random effects

ESF as a particular case.

Because no common parameter is assumed across the sub-models, the log-likelihood $l(\boldsymbol{\theta}_c)$ of each sub-model may be independently maximized to maximize $l(\boldsymbol{\theta}_{1:C}) = \sum_{c=1}^{C} l(\boldsymbol{\theta}_c)$. The resulting estimate/predictor are (see Appendix 2)

$$\begin{bmatrix} \hat{\mathbf{b}}_c \\ \hat{\mathbf{u}}_c \end{bmatrix} = \begin{bmatrix} \mathbf{X}'_c \ddot{\mathbf{W}}_c \mathbf{X}_c & \mathbf{X}'_c \ddot{\mathbf{W}}_c \mathbf{E}_c \mathbf{V}_c(\boldsymbol{\theta}_c) \\ \mathbf{V}_c(\boldsymbol{\theta}_c) \mathbf{E}'_c \ddot{\mathbf{W}}_c \mathbf{X}_c & \mathbf{V}_c(\boldsymbol{\theta}_c) \mathbf{E}'_c \ddot{\mathbf{W}}_c \mathbf{E}_c \mathbf{V}_c(\boldsymbol{\theta}_c) + \mathbf{I}_c \end{bmatrix}^{-1} \begin{bmatrix} \mathbf{X}'_c \ddot{\mathbf{W}}_c \mathbf{y}_c \\ \mathbf{V}_c(\boldsymbol{\theta}_c) \mathbf{E}'_c \ddot{\mathbf{W}}_c \mathbf{y}_c \end{bmatrix}. \quad (23)$$

The estimate of $\hat{\sigma}_c^2$ is also obtained by maximizing the likelihood, yielding

$$\hat{\sigma}_c^2 = \frac{\hat{\boldsymbol{\varepsilon}}'_c \ddot{\mathbf{W}}_c \hat{\boldsymbol{\varepsilon}}_c + \hat{\mathbf{u}}'_c \hat{\mathbf{u}}_c}{N_c - K}. \quad (24)$$

The aggregated SVC models in Eqs. (18)-(20) are estimated as follows:

(i) Partition a study area into $C$ - 1 spatial clusters and determine the model weights $w_c(s_i)$.

(ii) Estimate $C$ - 1 local sub-models by applying the following procedure for each $c$:

   (ii-1) Estimate $\boldsymbol{\theta}_c$ by maximizing $l(\boldsymbol{\theta}_c)$ (Eq. (22)), using subsamples satisfying $w_c(s_i) > 0$.

   (ii-2) Estimate/predict $\hat{\mathbf{b}}_c, \hat{\mathbf{u}}_c, \hat{\sigma}_c^2$ by substituting the estimated $\hat{\boldsymbol{\theta}}_c$ into Eq. (23) and (24).

   (ii-3) Predict the sub-model SVC using $\hat{\beta}_{k,c*}(s_i) = \hat{b}_{k,c} + \sum_{l=0}^{L_c} e_l(s_i) v_l(\hat{\boldsymbol{\theta}}_{k,c}) \hat{u}_{l,c}$ (see Eq. (20)).

(iii) Estimate the global sub-model with $L = 200$ using all samples in the same manner.

(iv) Predict the aggregated SVC using Eq. (19): $\hat{\beta}_{k*}(s_i) = \frac{1}{\sum_{c=1}^{C} w_c(s_i; \hat{\sigma}_c^2)} \sum_{c=1}^{C} w_c(s_i; \hat{\sigma}_c^2) \hat{\beta}_{k,c*}(s_i)$

(v) Evaluate the marginal likelihood $l(\boldsymbol{\theta}_{1:C})$ using Eq. (22).

For the clustering in step (i), we applied the *k*-means method because of the computational efficiency, considering the spatial coordinates.

Each sub-model has $3K + 1$ (free) parameters, and the aggregate model has $P_{1:C} = C(3K + 1)$ parameters. Likelihood-based model comparison and selection is available by considering the number of parameters. For example, the marginal Bayesian information criterion (BIC) (see Muller et al., 2013), which is given by $BIC = -2 \log l(\boldsymbol{\theta}_{1:C}) - 2N \log(P_{1:C})$, is a useful evaluation tool for this purpose.

The BIC is also available to select *C*. However, it requires an iteration involving the model estimation while varying *C* that is less suitable for large samples because of the computational cost. Instead, we determine *C* such that the average sample size per cluster becomes approximately 600, which reasonably balances the modeling accuracy and computational efficiency based on the preliminary exploratory analysis summarized in the Appendix 4.

Hereafter, we label to the proposed method ESF model aggregation (ESF-MA).

4.5. Salient ESF-MA property

ESF-MA has several advantages over ESF. First, unlike conventional ESF, which assumes a single model (i.e., $C = 1$), ESF-MA synthesizes $C$ sub-models, where a larger $C$ requires more parameters and eigenvectors. This property is helpful for adaptively determining model complexity by increasing $C$ according to the sample size. This property contrasts with the basic ESF, which has difficulty balancing accuracy and computational efficiency (see Section 2.2).

Second, procedure is practical. The aggregated model is easily constructed by synthesizing the outcomes from the sub-models that are independently estimated from each other. In addition, as explained above, likelihood-based model comparison is available, similar to the conventional spatial regression methods.

Third, it is computationally efficient. The sub-model estimation is easily parallelized. Besides, the fast likelihood maximization algorithm of Murakami and Griffith (2019b), which was developed for the random effects ESF model, is readily applicable to estimation the sub-model, which has the same form as the ESF model, to maximize $l(\boldsymbol{\theta}_c)$. Later, the algorithm is applied for all the sub-models. The fast eigen-decomposition of Murakami and Griffith (2019a) is also available to reduce the computational complexity $O(N_c^3)$ for extracting the $\mathbf{E}_c$ matrix to $O(N_c)$. This approximation is applied to the global sub-model assuming a large sample size $N_c$.

In summary, ESF-MA is expected to be an accurate, practical, and fast spatial modeling method that is suitable for large $N$. Section 5 verifies these advantages using simulation experiments.

5. Monte Carlo simulation experiments

This section investigates the modeling accuracy and computational efficiency using Monte Carlo experiments. Section 5.1 compares the possible specifications of the proposed model, and Section 5.2 reports comparisons with existing models.

5.1. Comparisons across specifications

This section examines whether the proposed method improves modeling accuracy relative to the conventional (random effects) ESF. In addition, it also scrutinizes how to specify the ESF-MA model through a comparison across possible specifications.

5.1.1. Experimental design

The models being compared are the conventional random effects ESF with $L = 200$ (ESF). For our model, we considered ESF-MA$_{L0}$, ESF-MA$_{GL0}$, ESF-MA$_{L}$, and ESF-MA$_{GL}$, as summarized in Table 2. The first two assume non-overlapping prior weights Eq. (9), while the latter two assume overlapping weights Eq. (10). ESF-MA$_{GL}$. ESF-MA$_{GL0}$, and ESF-MA$_{GL}$ aggregate the local and global sub-models, whereas the others only aggregate the local sub-models. All these models including ESF are estimated by applying the computationally efficient routine explained in Section 4.5.

Table 1: Specifications for ESF-MA.

|              | Global model | Local model | Overlap across clusters |
|--------------|:------------:|:-----------:|:-----------------------:|
| ESF-MA$_{L0}$  |              | ×           |                         |
| ESF-MA$_{GL0}$ | ×            | ×           |                         |
| ESF-MA$_{L}$   |              | ×           | ×                       |
| ESF-MA$_{GL}$  | ×            | ×           | ×                       |

These models are fitted on the synthetic data generated with

$$y(s_i) = \sum_{k=1}^{3} x_k(s_i)\beta_k(s_i) + \sum_{k=4}^{5} x_k(s_i)b_k + e(s_i), \quad e(s_i) \sim N(0,1), \quad (24)$$

$$\beta_k(s_i) = b_k + g_k(s_k), \quad g_k(s_k) \sim N(0, \tau_k^2 c(d_{i,j}; r)),$$

which assumes SVCs for $k \in \{1,2,3\}$ and constant coefficients $b_k$ for $k \in \{4,5\}$. The $k$-th SVC obeys a spatially dependent Gaussian process $g_k(s_k)$ with covariance $c(d_{i,j}; r) = \exp(-d_{i,j}/r)$ that exponentially decays as the distance $d_{i,j}$ increases. We assumed $[b_1, b_2, b_3, b_4, b_5] = [1, 2, -0.5, 1, -1]$ and $[\tau_1^2, \tau_2^2, \tau_3^2] = [1, 2, 1]$. Subsequently, we call $\beta_2(s_i)$, which has large variation ($\tau_2^2$), strong SVC and $\beta_3(s_i)$ weak SVC. The explanatory variables are generated with

$$x_k(s_i) = 0.5 g_{0,k}(s_i) + 0.5 g_{x,k}(s_i),$$
$$g_{0,k}(s_i) \sim N(0,1), \quad g_{x,k}(s_k) \sim N(0, c(d_{i,j}; 1))$$
(25)

which assumes moderate noise variation and spatial variation.

The data are generated on a $\sqrt{N} \times \sqrt{N}$ regularly spaced grid of points across a two-dimensional unit square. We assumed a sample size of $N \in \{50^2, 60^2, 70^2, 80^2, 100^2, 120^2\}$ and a range parameters $r \in \{0.5, 1.0, 2.0\}$. For each case, parameter estimation was performed 400 times, and the RMSE and computation time were compared across the models. The R package spmoran (version 0.2.2.9; https://cran.r-project.org/web/packages/spmoran/index.html) is used to estimate the ESF-based SVC model [Eqs. (1)-(2)]. Mac version 12.6.1 with 64 GB of memory (processor: 2.7 GHz 12 core Intel Xeon E5) is used for the calculations in Sections 5 and 6.

5.1.2. Simulation result

Figure 4 compares the RMSEs for the strong and weak SVCs ($\beta_2(s_i)$ and $\beta_3(s_i)$). In most cases, the four proposed models outperformed conventional ESF. Exceptionally, when $r = 2.0$, the weak SVCs estimated from ESF-MA$_{L0}$ and ESF-MA$_L$ exhibit larger RMSEs than the basic ESF. These local model-only aggregations perform poorly when the true process has a large-scale map pattern. By contrast, ESF-MA$_{GL0}$ and ESF-MA$_{GL}$ outperformed ESF across the inspected cases. The aggregation of local and global models is suggested to be helpful for accurately modeling strong and weak SVCs. ESF-MA$_{GL}$ has smaller RMSEs than ESF-MA$_{GL0}$ owing to the overlap in local clusters, mitigating the discontinuity of SVC processes near the cluster borders.

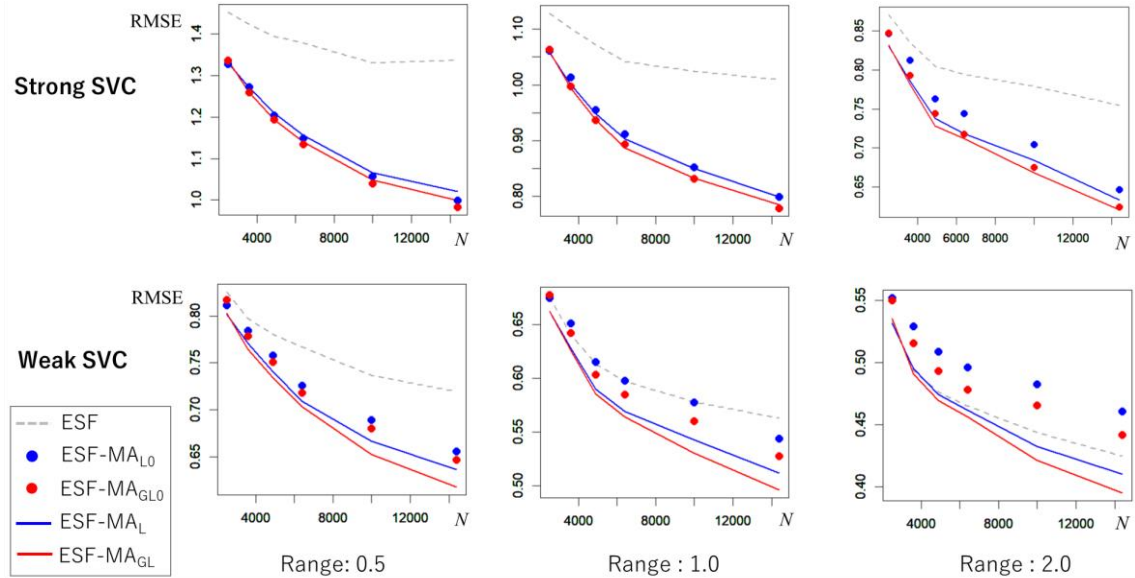

Figure 4: RMSE estimation accuracy for the SVCs.

Figure 5 compares the RMSEs for the constant coefficient $b_k$. All aggregation methods have smaller RMSE values than the conventional ESF. The proposed method is confirmed to be useful for not only SVC modeling, but also spatial regression estimating constant coefficients in the presence of residual spatial dependence (see LeSage and Pace, 2009).

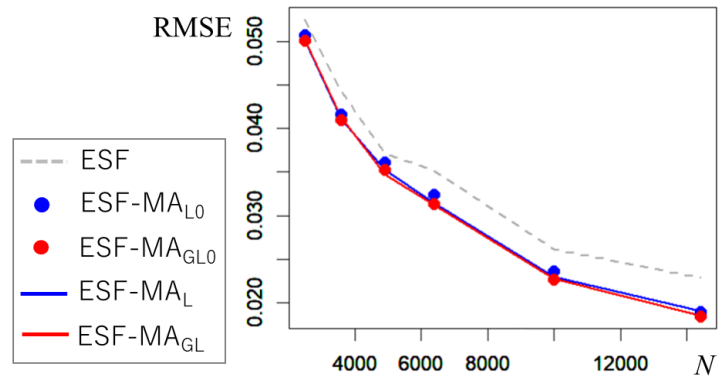

Figure 5: RMSE estimation accuracy for the constant coefficient.

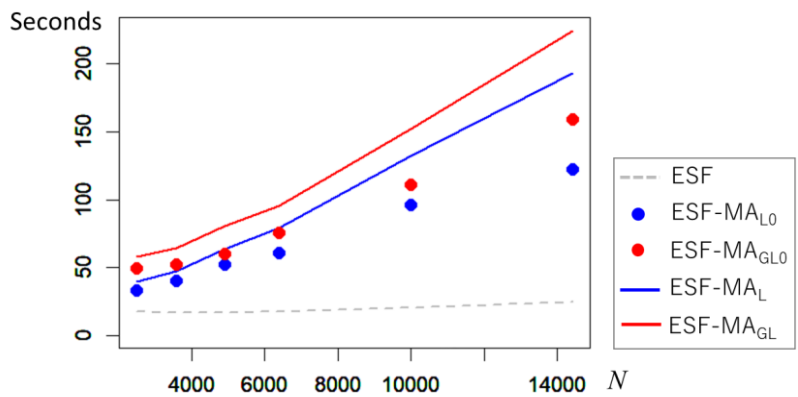

Figure 6. Average computation times of the ESF models without parallelization.

Finally, Figure 6 compares average computation time. ESF is quite fast due to the eigen-approximation and the efficient estimation algorithm implemented following Murakami and Griffith (2019a, b) (See Section 4.5). Since our aggregation methods apply the same algorithms to estimate each sub-model, they still exhibit reasonable computationally efficiency. Even if ESF-MA$_{GL}$, which is the most computationally intensive, the computation time increases only linearly with respect to the sample size, taking an average of only 224 seconds for 14,000 samples.

5.2. Comparison with GWR

This section compares the accuracy and computational time for ESF-MA$_{GL}$, which was the most accurate in Section 5.1, with GWR, which is a popular SVC modeling approach.

5.2.1. Experimental design

Following the assumption in GWR, we generated synthetic samples with the following model without a constant coefficient:

$$y(s_i) = \sum_{k=1}^{3} x_k(s_i)\beta_k(s_i) + e(s_i), \quad e(s_i) \sim N(0,1), \tag{26}$$

$$\beta_k(s_i) = b_k + g_k(s_k), \quad g_k(s_k) \sim N(0, \tau_k^2 c(d_{i,j}; r)).$$

The other settings are the same as reported in the previous section. The GWmodel package (version 2.2-9; https://cran.r-project.org/web/packages/GWmodel/index.html; Lu et al., 2014) is used to estimate the GWR.

5.2.2. Result

Figure 7 plots strong SVCs $\beta_2(s_i)$ estimated when $N = 100^2$. As expected, ESF tends to have excessively smoothed map patterns, whereas ESF-MA$_{GL}$ does not. The SVC estimates of ESF-MA$_{GL}$ are visually similar to those for GWR, which does not suffer from the degeneracy problem.

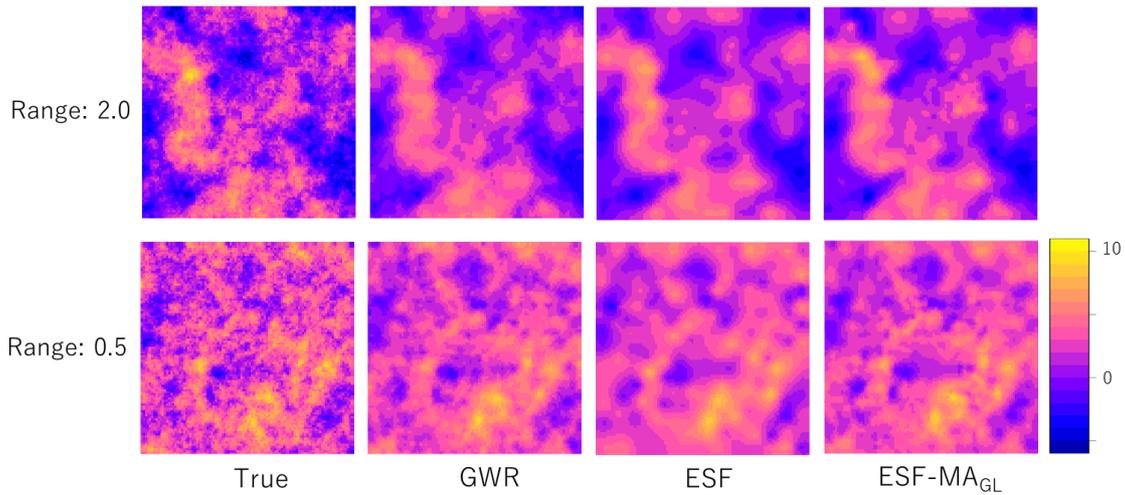

Figure 7: Estimated strong SVCs ($\beta_2(s_i)$) in scenarios with $N = 100^2$ and $r \in \{0.5, 2.0\}$.

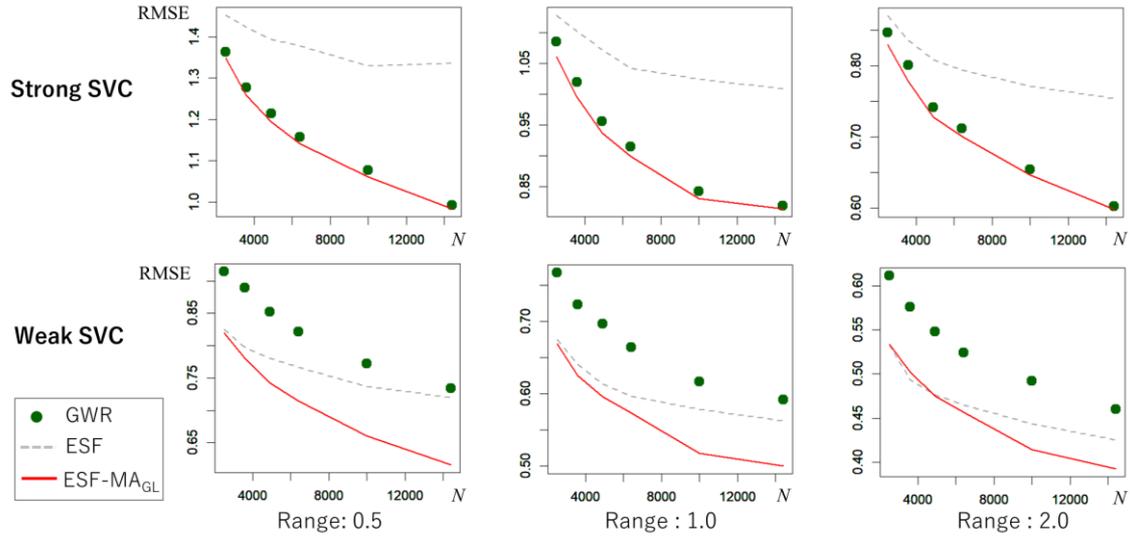

Figure 8: Comparison of SVC RMSE estimation accuracy with conventional methods.

The RMSE values are compared in Figure 8. the ESF has a large RMSE for strong SVCs, which is attributable to the degeneracy problem limiting model flexibility. Improvement in the ESF model accuracy with increasing $N$ is relatively slow due to this problem. In contrast, GWR had a large RMSE for weak SVCs, which is because it only considers local samples even if more samples are required to detect weak spatial patterns.

The proposed ESF-MA$_{GL}$ achieved the smallest RMSE values in all cases. A strong SVC attained much smaller RMSE values than those of ESF by introducing local sub-models, mitigating the degeneracy problem. Weak SVCs have smaller RMSEs than GWR possibly owing to the global sub-model that helps stabilize the estimation of weak patterns. Considering the improvement in accuracy with increasing $N$, ESF-MA$_{GL}$ is more suitable for large samples than the conventional ESF.

Finally, Figure 9 compares the mean computation time of five trials for each $N \in$ {4900,10000,19600,40000} assuming $r = 1.0$. GWR and ESF-MA$_{GL}$ were parallelized on a 12 core Mac computer and compared with those without parallelization.

Figure 9 suggests that the proposed method is applicable to large samples even without parallelization. The non-parallelized ESF-MA$_{GL}$ took 680.9 seconds on average for 40,000 samples. The result also indicates that the computation time is significantly shortened by parallelization. In particular, the increase in the computation time with respect to $N$ was considerably smaller than that for the parallelized GWR. For example, the average computation times were 103.8 seconds for $N$ = 4,900 and 232.9 seconds for $N$ =40,000. Parallelized ESF-MA$_{GL}$ appears to be helpful in modeling very large samples.

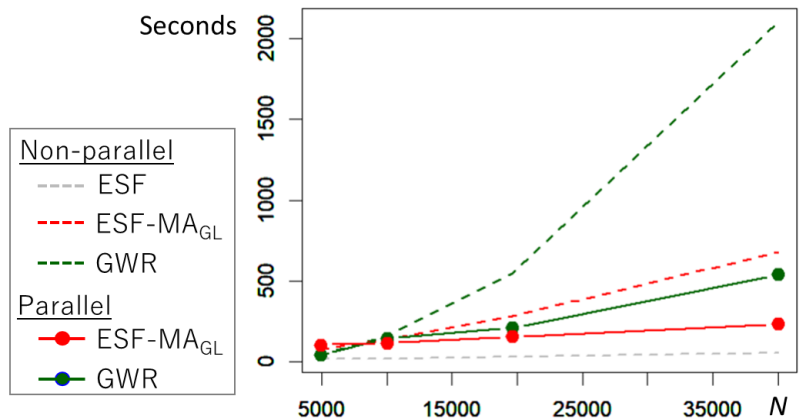

Figure 9: Average computation times for ESF and GWR models. 12 core CPU is used for the parallelization.

## 5.3. Comparison with multiscale GWR (MGWR) model

MGWR (Yang, 2014; Fotheringham et al., 2017) is an extension of GWR to estimate the spatial scale for each SVC by assigning a bandwidth to each. After Murakami et al. (2019) and Shabrina et al. (2019) confirmed the accuracy of MGWR, it became recognized as a flexible alternative of GWR (see Comber et al., 2022). Given that outcome, this section compares ESF-MA$_{GL}$ with MGWR.

### 5.3.1. Experimental design

Following MGWR, synthetic samples are generated from the SVC model Eq. (26) whose range parameter $r$ for $k \in \{1,2,3\}$ is replaced with $\{r_1, r_2, r_3\} = \{1.0, 0.5, 2.0\}$ assuming moderate-, small-, and large-scale map patterns for $\beta_1(s_i), \beta_2(s_i), \beta_3(s_i)$. Considering the computational cost of MGWR and the suitability of ESF-MA$_{GL}$ for large samples, the sample size is $N = 2500$. The other settings are the same as for the previous section.

### 5.3.2. Results

Because the difference between MGWR and ESF-MA$_{GL}$ is relatively small, the comparison results are summarized in Table 3. While MGWR and ESF-MA$_{GL}$ indicated similar RMSE and mean absolute error (MAE) values, the latter indicates better performance across cases. The computation time of ESF-MA$_{GL}$ is much shorter than MGWR: on average, ESF-MA$_{GL}$ is about 60.57 (=3,723/61.47) times faster than MGWR. Considering the linear computation time with respect to $N$ shown in Fig. 8, the difference should be even greater for large samples.[4]

Table 3: RMSE, MAE, and computation time of MGWR and ESF-MA$_{GL}$. Bold indicates better performance. 1Q. and 3Q. stand for the 1st and 3rd quantile.

|  | RMSE | | | | | | Computation time (seconds) | |
|---|---|---|---|---|---|---|---|---|
|  | $\beta_1(s_i)$ | | $\beta_2(s_i)$ | | $\beta_3(s_i)$ | | | |
|  | MGWR | ESF-MA$_{GL}$ | MGWR | ESF-MA$_{GL}$ | MGWR | ESF-MA$_{GL}$ | MGWR | ESF-MA$_{GL}$ |
| Min. | 0.512 | **0.509** | 0.698 | **0.692** | 0.679 | **0.664** | 2,870 | **49.38** |
| 1Q. | 0.573 | **0.562** | 0.757 | **0.757** | 0.814 | **0.810** | 3,405 | **58.78** |
| Median | 0.590 | **0.576** | 0.777 | **0.773** | 0.865 | **0.841** | 3,685 | **61.10** |
| Mean | 0.593 | **0.578** | 0.776 | **0.772** | 0.866 | **0.851** | 3,723 | **61.47** |
| 3Q. | 0.616 | **0.594** | 0.793 | **0.787** | 0.919 | **0.913** | 4,099 | **64.38** |
| Max. | 0.669 | **0.656** | 0.835 | **0.832** | 1.036 | **1.018** | 4,660 | **81.85** |
|  | MAE | | | | | | | |
|  | MGWR | ESF-MA$_{GL}$ | MGWR | ESF-MA$_{GL}$ | MGWR | ESF-MA$_{GL}$ | | |
| Min. | 0.407 | **0.404** | 0.552 | **0.549** | 0.535 | **0.526** | | |
| 1Q. | 0.456 | **0.446** | 0.601 | **0.600** | 0.645 | **0.640** | | |
| Median | 0.470 | **0.457** | 0.618 | **0.615** | 0.686 | **0.666** | | |
| Mean | 0.471 | **0.460** | 0.618 | **0.615** | 0.687 | **0.677** | | |
| 3Q. | 0.489 | **0.471** | 0.631 | **0.627** | 0.731 | **0.730** | | |
| Max. | 0.522 | **0.519** | 0.664 | **0.660** | 0.831 | **0.813** | | |

---

[4] The MGWR estimation requires an iterative process whose computational complexity order is $KN^2\log(N)$, until convergence (Li and Fotheringham, 2020).

In summary, Section 5 corroborates that the proposed method is a fast and accurate alternative to conventional ESF, GWR, or MGWR for large samples. The next section demonstrates its usefulness through an empirical land price analysis case study.

6. An empirical application

This section applies ESF-MA to an empirical land price analysis to demonstrate its usefulness for real-world data.

6.1. Case study overview

In Japan, where the population is continues to decline, there is a need to concentrate urban facilities and habitat areas along public transportation networks to reduce infrastructure maintenance costs and carbon emissions while increasing livability and environmental friendliness (see Yamagata and Seya, 2013). As a first step toward such sustainable development, we evaluated the influence of train and bus networks on urban structures by analyzing officially assessed residential land price data in 2010 in Japan[5] ($N$ = 41,266; source: National Land Numerical Information (NLNI) download service (https://nlftp.mlit.go.jp/ksj/)). As plotted in Figure 10 (a), land prices increased around Tokyo, Osaka, and other major cities. Train and bus networks are densely spread throughout these cities. Figure 10 (b) shows the training network.

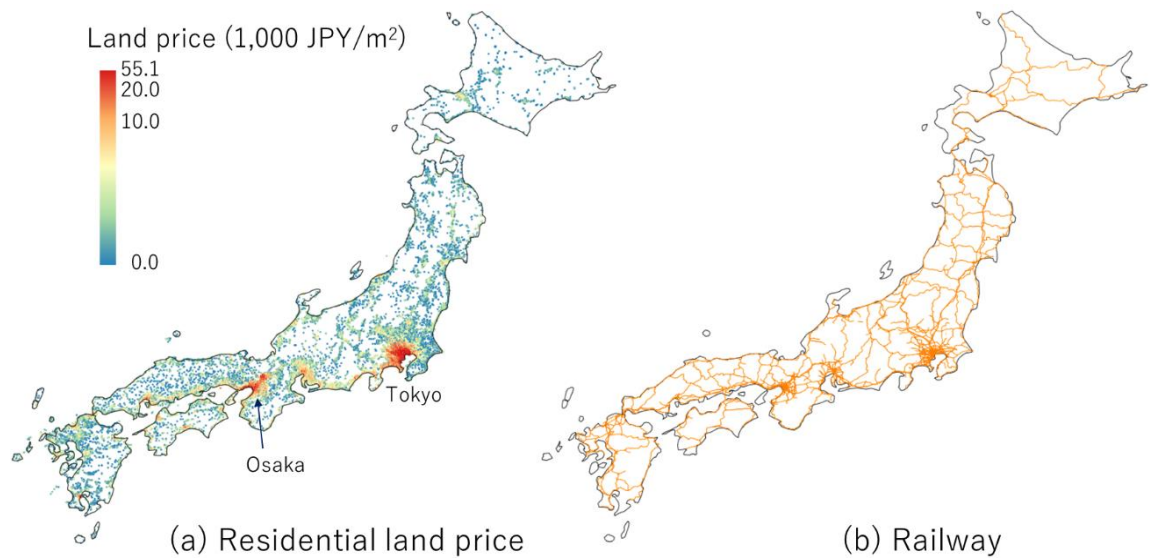

Figure 10: Map of (a) residential land price and (b) passenger railway network in 2010.

---

[5] The Okinawa prefecture is excluded from this study area because it is located far from the main island of Japan, and has no rail network.

This section applies the ESF-based SVC model and ESF-MA to this analysis. The response variable is log-land prices (JPY/km$^2$). The explanatory variables are the following: Euclidean distance to the nearest railway station, in kilometers (Train); Euclidean distance to the nearest bus station, in kilometers (Bus); dummy variable denoting commercial area by 1, and 0 otherwise (Commerce); dummy variable denoting industrial area by 1, and 0 otherwise (Industry); and, estimated flood inundation depth in meters (Flood). We assumed SVCs for the Train and Bus variates, which are our primary attributes of interests, and constant coefficients were assumed for the other three. For ESF-MA, based on the experiments in Section 5 and Appendix. 4, we used 69 spatial clusters extracted using the *k*-means method assuming approximately 600 samples per cluster (Figure 11).

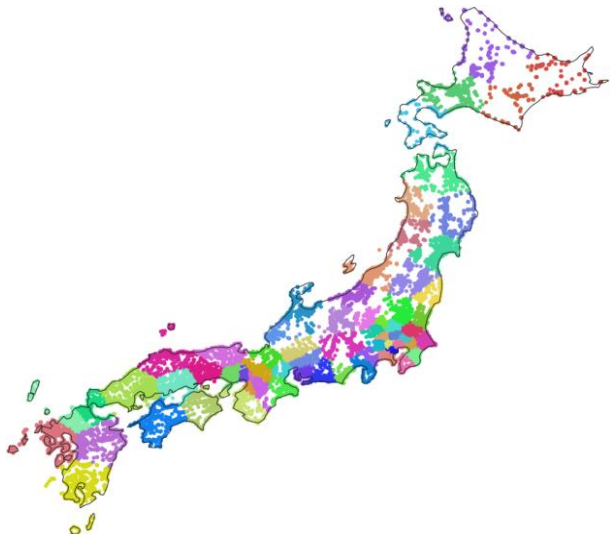

Figure 11: Spatial clusters assumed when estimating ESF-MA.

6.2. Case study result

Figures 12 plots the estimated SVCs. Spatially varying intercepts estimated from the two models indicate high land prices nearby Tokyo, Osaka, and other major cities. The Train and Bus variable SVCs estimated with ESF-MA tend to have local map patterns while those estimated with ESF tend to be over smooth. Overall, the coefficients estimated with ESF-MA are more interpretable than conventional ESF ones, as we further explain below.

Based on ESF-MA, variate Train exhibits pronounced covariation with land prices in the greater Tokyo and Osaka metropolitan areas. The covariation is also strong along railways near other major cities, including Fukuoka, Niigata, and Fukushima. These results are reasonable because trains are widely used for commuting in major cities. Meanwhile, variate Bus network displays prominent covariation with land prices in Hokkaido where many train routes have been converted to bus routes over the years; the total length of passenger railway routes is 5,320 km in 1980, down to 2,481 km as of 2010 (source: NLNI). A bus network has been suggested as an important covariate of urbanization in this region. It also appears to be a conspicuous covariate in the Southern Kyushu region, where the train network is sparse (Figure 10). In summary, the Train variable strongly covariates with land prices around major cities along railways, whereas the Bus variable strongly covariates with land prices in less urbanized areas with limited rail accessibility.

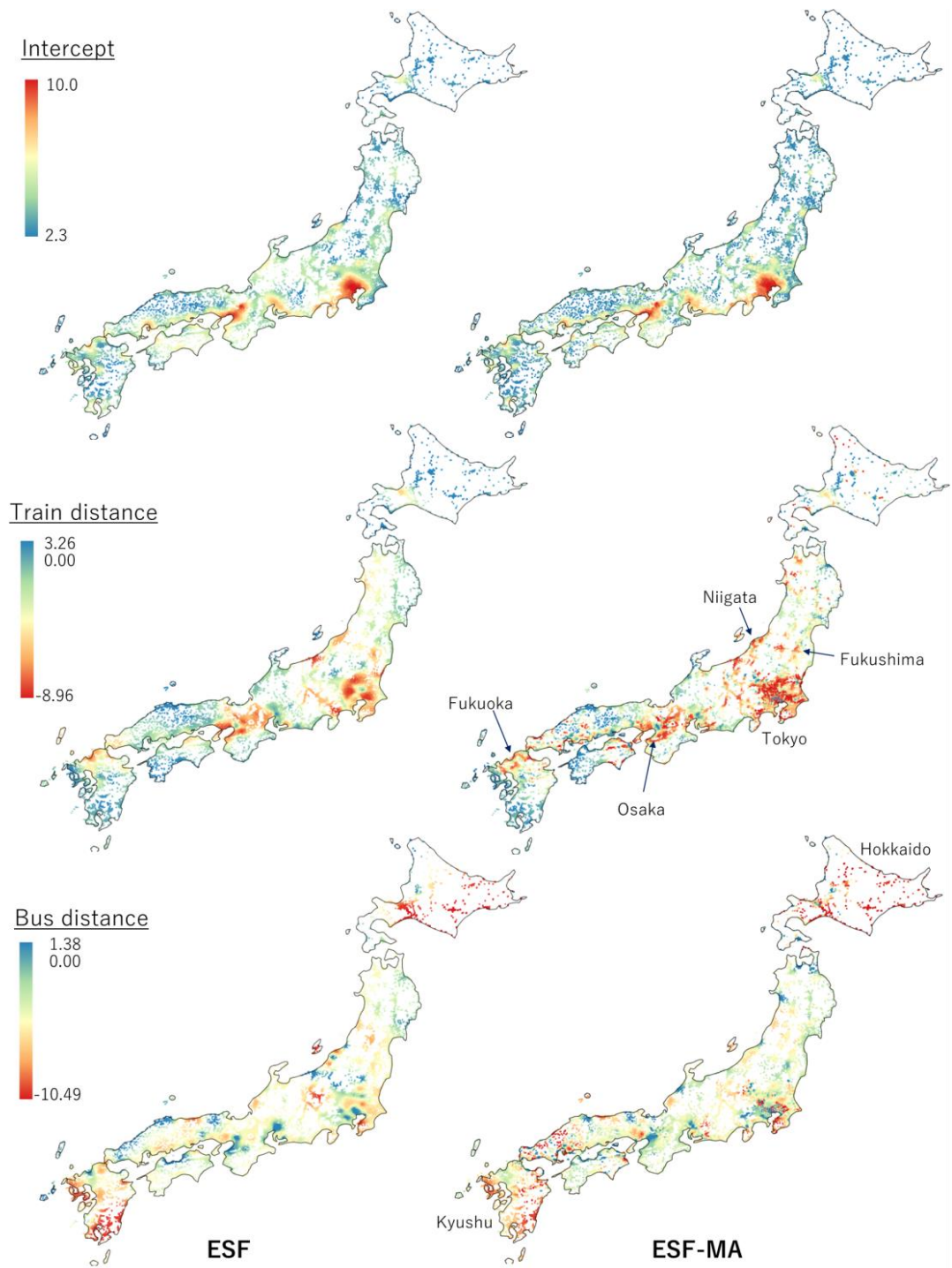

Figure 12: SVCs estimates from the conventional ESF.

Figures 14 compares the estimated relationship between the Train variable and the log-land prices by the eight regions shown in Figure 13. Figure 15 displays the same for the Bus variable. As shown in Figure 14, land prices rapidly decline as distance to the nearest train station increases in the Tokyo and Osaka areas. This result confirms the importance of a train network in major metropolitan areas. In contrast, land prices quickly decline as the distance to bus stops increase in Hokkaido, although less so in other regions. This case study verifies that ESF-MA can be useful for uncovering regional heterogeneity, partly because of its framework for modeling local heterogeneity utilizing local sub-models.

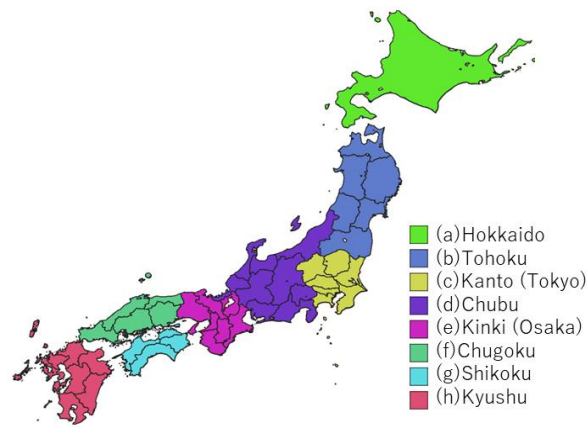

Figure 13: Prefectures in Japan (except for Okinawa) colored by regional classifications (a) – (h).

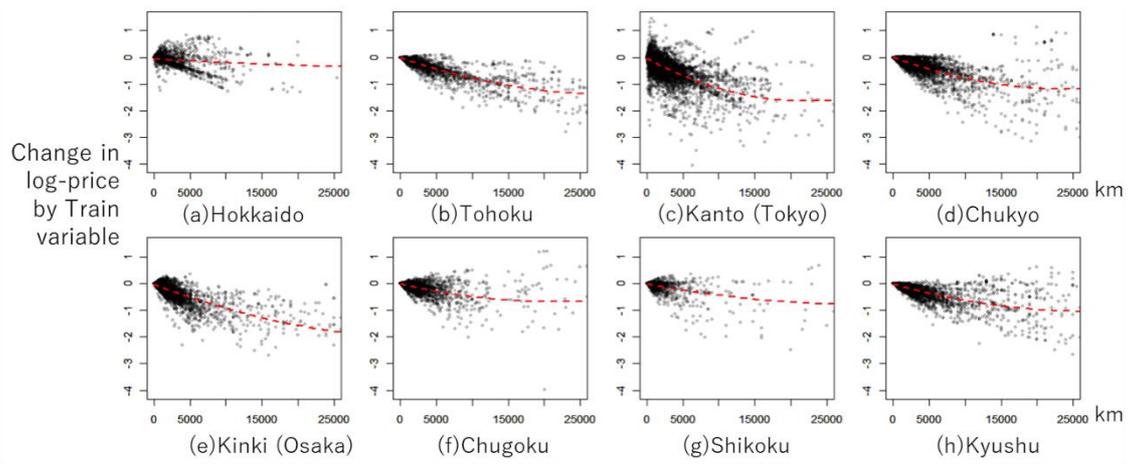

Figure 14: Estimated Train variable relationships by region (x-axis: distance to train station $x_{Train}(s_i)$; y-axis: estimated associations measured by $x_{Train}(s_i)\hat{\beta}_{Train}(s_i)$). Fitted quadratic polynomials are denoted by red dashed trend lines.

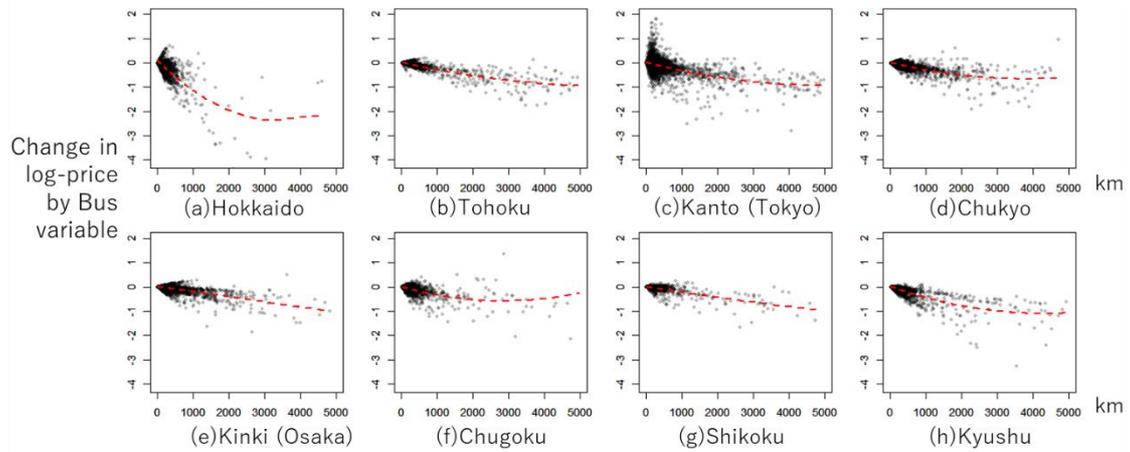

Figure 15: Estimated Bus variable relationships by region (x-axis: distance to bus station $x_{Bus}(s_i)$; y-axis: estimated associations measured by $x_{Bus}(s_i)\hat{\beta}_{Bus}(s_i)$). Fitted quadratic polynomials are denoted by red dashed trend lines.

Table 4 summarizes the estimated constant coefficients as well as marginal log-likelihood. The greater log-likelihood value for ESF-MA confirms its superior accuracy. ESF and ESF-MA estimated statistically significant positive associations for Commerce, and negative associations for Industry. The former reflects convenience in commercial areas, whereas the latter reflects the unattractiveness of industrial areas, most likely because of such negative spatial externalities as air pollution and noise. Flood is positively significant in the ESF and insignificant in the ESF-MA specification. Based on our model, flood risk is not appropriately reflected as a negative value in residential land prices. Considering projected increasing future flood risk (see Hirabayashi et al., 2013), residential land regulation might be required to reduce people living in areas which are predicted flood-prone in the future.

Table 4: Estimated constant coefficients and marginal log-likelihood values.

|  | ESF | | | ESF-MA | | |
| --- | --- | --- | --- | --- | --- | --- |
|  | Estimates | t-value | Signif. | Estimates | t-value | Signif. |
| Commerce | 0.595 | 49.113 | ***[1] | 0.448 | 25.662 | *** |
| Industry | -0.073 | -3.297 | *** | -0.227 | -7.131 | *** |
| Flood | 0.004 | 3.011 | *** | 0.001 | 0.597 |  |
| Log-likelihood | -47,055 | | | -37,721 | | |

[1] *** represents the statistical significance at the 1 % level.

7. Concluding remarks

We developed a spatial regression method for aggregating/averaging local and global ESF models. Unlike the conventional ESF, which suffers from a degeneracy problem, the proposed ESF-MA is scalable for large samples in terms of accuracy and computational efficiency. This empirical case study included in this paper demonstrates that the proposed method is useful to enhance interpretability of SVC estimates. ESF-MA would be a sensible choice for large-scale spatial modeling.

The ESF-MA is easily adapted to specific purposes. For example, sub-models may be defined by prefecture to analyze heterogeneity across prefectures. Sub-models at country, prefecture, and city levels also can be useful for multiscale analyses of, for example, COVID-19 infections and socioeconomic phenomena (Griffith and Li, 2021; Griffith, 2022). In addition, sub-sample determination is possible considering not only spatial coordinates, but also other features. For example, sub-models by income level might be helpful to flexibly analyze regional income considering map patterns varying depending on income levels. Sub-models distributed over space and time might be useful for scalable spatiotemporal modeling (see Nishi et al., 2022).

Unlike most spatial regression techniques that assume a single model, ESF-MA

constructs a flexible model by aggregating multiple sub-models. Such ensemble learning methods have been actively studied in the machine learning literature (Sagi and Rokach, 2018). Further extension of ESF-MA by incorporating machine learning techniques would be important for improving scalability and applicability to prediction, multitask learning, transfer learning, and other learning tasks, which increasingly are attracting attention in geographic information science (see Mai et al., 2022; Seya and Shiroi. 2022).

It is also possible to employ other aggregation methods to extend the ESF-MA. Although ESF-MA considers the error variance for weighting sub-models, Rullière et al. (2016) and Nakai-Kasai and Tanaka (2022) consider the covariance among sub-models for such weighting. The covariance-based weighting improves modeling accuracy by down-weighting redundant sub-models that overlap with other sub-models. Other aggregation methods, which might be helpful to extend ESF-MA, have been developed (e.g., Park and Apley, 2018; and Edwards and Gramacy, 2021) as reviewed in Liu et al. (2020).

Cluster optimization is another issue. Li and Sang (2019) and Sugasawa and Murakami (2021) simultaneously optimize clusters and regression models to estimate SVCs. Gramacy and Lee (2008) and Van Stein et al. (2020) apply tree-based partitioning to determine the clusters.

The other way to enhance flexibility is to aggregate sub-models with other statistical models. For example, aggregation/averaging with a linear additive/mixed model (e.g., Wood, 2017), which includes the (RE)ESF model as a special case, would be useful to account for temporal, nonlinear, and group-wise effects. Boosting-like extensions generating sub-models randomly and aggregating them sequentially to maximize model accuracy, is also possible (Tautvaišas and Ilinskas, 2022).

In conclusion, there are many directions for extending the proposed method for complex and large-scale spatial and spatiotemporal modeling efforts. Extension of ESF-MA is an important topic for making the most of large data in regional science, environmental science, epidemiology, and other disciplines.

The proposed method is implemented in the addlearn_local function in the spmoran package version 0.3.0 or later.

Appendix 1: Derivation of the aggregated model

In Eq. (17), our model was defined as

$$p(\mathbf{y}, \mathbf{u}_{1:C} | \mathbf{b}_{1:C}, \boldsymbol{\theta}_{1:C}, \sigma^2_{1:C}) \propto p(\mathbf{y} | \mathbf{u}_{1:C}, \mathbf{b}_{1:C}, \boldsymbol{\theta}_{1:C}, \sigma^2_{1:C}) p(\mathbf{u}_{1:C} | \sigma^2_{1:C}), \tag{A1}$$

where

$$p(\mathbf{y} | \mathbf{u}_{1:C}, \mathbf{b}_{1:C}, \boldsymbol{\theta}_{1:C}, \sigma^2_{1:C}) = \prod_{i_c=1}^{N_c} p(y(s_i) | \mathbf{u}_{1:C}, \mathbf{b}_{1:C}, \boldsymbol{\theta}_{1:C}, \sigma^2_{1:C}), \tag{A2}$$

which is the product of the following data model

$$p(y(s_i) | \mathbf{u}_{1:C}, \mathbf{b}_{1:C}, \boldsymbol{\theta}_{1:C}, \sigma^2_{1:C}) = \prod_{c=1}^{C} p(y(s_i) | \mathbf{u}_c, \mathbf{b}_c, \boldsymbol{\theta}_c, \sigma^2_c)^{w_c(s_i)}, \tag{A3}$$

which is identical to the usual generalized PoE model (Eq. 4). Eq. (A3) is expressed following Eqs. (6-8) as

$$y(s_i) | \mathbf{u}_{1:C}, \mathbf{b}_{1:C}, \boldsymbol{\theta}_{1:C}, \sigma^2_{1:C} \sim N(\mu_*(s_i), \sigma^2_*(s_i)), \tag{A4}$$

where $\mu_*(s_i) = \frac{\sum_c w_c(s_i; \sigma^2_c) \mu_c(s_i)}{\sum_c w_c(s_i; \sigma^2_c)}$ and $\sigma^2_*(s_i) = \left( \sum_{c=1}^{C} \frac{w_c(s_i)}{\sigma^2_c} \right)^{-1}$. By substituting $\mu_c(s_i) = \sum_{k=0}^{K} x_k(s_i) \left( b_{k,c} + \sum_{l=0}^{L_c} e_l(s_i) v_l(\boldsymbol{\theta}_{k,c}) u_{l,c} \right)$ which is assumed in Eqs. (11-12) into Eq. (A4), and organizing these equations, we have

$$y(s_i) | M_{1:c} \sim N \left( \sum_{k=1}^{K} x_k(s_i) \beta_k(s_i), \sigma^2(s_i) \right), \tag{A5}$$

$$\beta_k(s_i) = \frac{1}{\sum_{c=1}^{C} w_c(s_i; \sigma^2_c)} \sum_{c=1}^{C} w_c(s_i; \sigma^2_c) \left( b_{k,c} + \sum_{l=0}^{L_c} e_l(s_i) v_l(\boldsymbol{\theta}_{k,c}) u_{l,c} \right), \tag{A6}$$

where $y(s_i) | \mathbf{u}_{1:C}, \mathbf{b}_{1:C}, \boldsymbol{\theta}_{1:C}, \sigma^2_{1:C}$ is replaced with $y(s_i) | M_{1:c}$. Regarding $\mathbf{u}_c$, we assumed it to obey a distribution $\frac{1}{Q} p(\mathbf{u}_c | \sigma^2_c)^{W_c/N_c}$ where $p(\mathbf{u}_c | \sigma^2_c)$ has the probability density function (PDF) of $N(\mathbf{0}, \sigma^2_c \mathbf{I}_c)$ and $Q$ is a normalizing constant. It is immediate to show that the distribution becomes $\mathbf{u}_c \sim N\left(\mathbf{0}, \frac{N_c \sigma^2_c}{W_c} \mathbf{I}_c \right)$. By substituting it into Eqs. (A5-A6), we obtain our model Eqs. (18-20).

Appendix 2: Derivation of the log-likelihood: $l(\boldsymbol{\theta}_{1:C}, \mathbf{b}_{1:C}) = \log L(\boldsymbol{\theta}_{1:C}, \mathbf{b}_{1:C})$

The $c$-th model has parameters $\{\mathbf{b}_c, \sigma^2_c\}$ and $\boldsymbol{\theta}_c = [\alpha_{c,1}, \ldots, \alpha_{c,L}, \tau^2_{c,1}, \ldots, \tau^2_{c,L}]'$. We assumed the following distribution for the explained variables in Eq. (15):

$$p(\mathbf{y} | \mathbf{u}_{1:C}, \mathbf{b}_{1:C}, \sigma^2_{1:C}, \boldsymbol{\theta}_{1:C}) = \prod_{c=1}^{C} \prod_{i=1}^{N} p(y(s_i) | \mathbf{u}_c, \mathbf{b}_c, \sigma^2_c, \boldsymbol{\theta}_c)^{w_c(s_i)} \tag{A7}$$

The equality in Eq. (A7) holds since $\sum_{c=1}^{C} w_c(s_{i_c}) = 1$. By noting that $p(y(s_i) | \mathbf{u}_c, \mathbf{b}_c, \sigma^2_c, \boldsymbol{\theta}_c)^0 = 1$, Eq. (A7) is rewritten as:

$$p(\mathbf{y}|\mathbf{u}_{1:C},\mathbf{b}_{1:C},\sigma^2_{1:C},\boldsymbol{\theta}_{1:C}) = \prod_{c=1}^{C}\prod_{i_c=1}^{N_c} p(y(s_{i_c})|\mathbf{u}_c,\mathbf{b}_c,\sigma^2_c,\boldsymbol{\theta}_c)^{w_c(s_{i_c})}, \tag{A8}$$

where

$$\prod_{i_c=1}^{N_c} p(y(s_{i_c})|\mathbf{u}_c,\mathbf{b}_c,\sigma^2_c,\boldsymbol{\theta}_c)^{w_c(s_{i_c})} = \frac{1}{(2\pi\sigma^2_c)^{W_c/2}} \exp\left(-\frac{\boldsymbol{\varepsilon}_c'\mathbf{W}_c\boldsymbol{\varepsilon}_c}{2\sigma^2_c}\right), \tag{A9}$$

and $\boldsymbol{\varepsilon}_c = \mathbf{y}_c - \mathbf{X}_c\mathbf{b}_c - \mathbf{E}_c\mathbf{V}_c(\boldsymbol{\theta}_c)\mathbf{u}_c$.

For $\mathbf{u}_c$, let us assume the following structure:

$$p(\mathbf{u}_c|\sigma^2_c) \propto \prod_{i_c=1}^{N_c} p_0(\mathbf{u}_c|\sigma^2_c)^{\frac{w_c(s_{i_c})}{N_c}} = p_0(\mathbf{u}_c|\sigma^2_c)^{z_c}, \tag{A10}$$

where $z_c = \frac{\sum_{i_c}^{N_c} w_c(s_{i_c})}{N_c} = \frac{W_c}{N_c}$ and $p_0(\mathbf{u}_c|\sigma^2_c)$ is a Gaussian PDF. The following equality holds

$$\begin{aligned}\int p_0(\mathbf{u}_c|\sigma^2_c)^{z_c} d\mathbf{u}_c &= \int \frac{1}{(2\pi\sigma^2_c)^{z_c L/2}} \exp\left(-\frac{z_c}{2\sigma^2_c}\mathbf{u}'_c\mathbf{u}_c\right) d\mathbf{u}_c \\ &= \frac{(2\pi\sigma^2_c)^{L/2}}{z_c^{L/2}(2\pi\sigma^2_c)^{z_c L/2}} \int \frac{z_c^{L/2}}{(2\pi\sigma^2_c)^{L/2}} \exp\left(-\frac{z_c}{2\sigma^2_c}\mathbf{u}'_c\mathbf{u}_c\right) d\mathbf{u}_c \\ &= \frac{(2\pi\sigma^2_c)^{L/2}}{z_c^{L/2}(2\pi\sigma^2_c)^{z_c L/2}}\end{aligned} \tag{A11}$$

where the integral part disappeared because this part is a Gaussian PDF whose integral equals one. Using Eq. (A11), $p(\mathbf{u}_c|\sigma^2_c)$ is expressed as

$$p(\mathbf{u}_c|\sigma^2_c) = \frac{p_0(\mathbf{u}_c|\sigma^2_c)^{z_c}}{\int p_0(\mathbf{u}_c|\sigma^2_c)^{z_c} d\mathbf{u}_c} = \frac{z_c^{L/2}}{(2\pi\sigma^2_c)^{L/2}} \exp\left(-\frac{z_c}{2\sigma^2_c}\mathbf{u}'_c\mathbf{u}_c\right), \tag{A12}$$

which is identical to the Gaussian PDF of $\mathbf{u}_c \sim N(\mathbf{0}_c, \sigma^2_c/z_c \mathbf{I}_c)$, meaning that $u_l \sim N\left(0, \frac{N_c}{W_c}\sigma^2_c\right)$ which we assumed in Eq. (20). By multiplying Eqs. (A8-A9) and (A12). the joint distribution $p(\mathbf{y},\mathbf{u}_{1:C}|\mathbf{b}_{1:C},\sigma^2_{1:C},\boldsymbol{\theta}_{1:C})$ is expressed as

$$\begin{aligned}p(\mathbf{y},\mathbf{u}_{1:C}|\mathbf{b}_{1:C},\sigma^2_{1:C},\boldsymbol{\theta}_{1:C}) &\propto f(\mathbf{y},\mathbf{u}_{1:C}|\mathbf{b}_{1:C},\sigma^2_{1:C},\boldsymbol{\theta}_{1:C}) = \prod_{c=1}^{C} f(\mathbf{y}_c,\mathbf{u}_c|\mathbf{b}_c,\sigma^2_c,\boldsymbol{\theta}_c), \\ f(\mathbf{y}_c,\mathbf{u}_c|\mathbf{b}_c,\sigma^2_c,\boldsymbol{\theta}_c) &= \frac{z_c^{L/2}}{(2\pi\sigma^2_c)^{(W_c+L)/2}} \exp\left(-\frac{\boldsymbol{\varepsilon}_c'\mathbf{W}_c\boldsymbol{\varepsilon}_c + z_c\mathbf{u}'_c\mathbf{u}_c}{2\sigma^2_c}\right),\end{aligned} \tag{A13}$$

$L(\mathbf{b}_{1:C},\sigma^2_{1:C},\boldsymbol{\theta}_{1:C}) = \int f(\mathbf{y},\mathbf{u}_{1:C}|\mathbf{b}_{1:C},\sigma^2_{1:C},\boldsymbol{\theta}_{1:C}) d\mathbf{u}_{1:C}$ is the marginal likelihood we wish to evaluate (see Bates, 2010).

We first consider deriving a closed-form expression of $L(\mathbf{b}_c,\sigma^2_c,\boldsymbol{\theta}_c) = \int f(\mathbf{y}_c,\mathbf{u}_c|\mathbf{b}_c,\sigma^2_c,\boldsymbol{\theta}_c) d\mathbf{u}_c$. We use the following equality, which is obtained through Taylor expansion[6]:

---

[6] Through Taylor expansion about $\hat{\mathbf{u}}_c$, we have $\log f(\mathbf{y}_c,\mathbf{u}_c|\mathbf{b}_c,\sigma^2_c,\boldsymbol{\theta}_c) = \log f(\mathbf{y}_c,\hat{\mathbf{u}}_c|\mathbf{b}_c,\sigma^2_c,\boldsymbol{\theta}_c) + \frac{1}{2}(\mathbf{u}_c -$

$$\boldsymbol{\varepsilon}_c'\mathbf{W}_c\boldsymbol{\varepsilon}_c + z_c\mathbf{u}'_c\mathbf{u}_c = z_c\big(\boldsymbol{\varepsilon}'_c\ddot{\mathbf{W}}_c\boldsymbol{\varepsilon}_c + \mathbf{u}'_c\mathbf{u}_c\big)$$
$$= z_c\left(\hat{\boldsymbol{\varepsilon}}'_c\ddot{\mathbf{W}}_c\hat{\boldsymbol{\varepsilon}}_c + \hat{\mathbf{u}}'_c\hat{\mathbf{u}}_c + (\mathbf{u}_c - \hat{\mathbf{u}}_c)'\mathbf{M}_c(\mathbf{u}_c - \hat{\mathbf{u}}_c)\right), \quad (A14)$$

where $\ddot{\mathbf{W}}_c = z_c^{-1}\mathbf{W}_c$, $\mathbf{M}_c = \boldsymbol{\Lambda}_c(\boldsymbol{\theta}_c)\mathbf{E}'_c\ddot{\mathbf{W}}_c\mathbf{E}_c\boldsymbol{\Lambda}_c(\boldsymbol{\theta}_c) + \mathbf{I}_c$, $\hat{\boldsymbol{\varepsilon}}_c = \mathbf{y}_c - \mathbf{X}_c\mathbf{b}_c - \mathbf{E}_c\mathbf{V}_c(\hat{\boldsymbol{\theta}}_c)\hat{\mathbf{u}}_c$. By using Eqs (A8) and (A9), the likelihood for the $c$-th sub-model is expanded as

$$L(\sigma_c^2, \boldsymbol{\theta}_c, \mathbf{b}_c) = \int f(\mathbf{y}_c, \mathbf{u}_c | \mathbf{b}_c, \sigma_c^2, \boldsymbol{\theta}_c)\, \mathbf{u}_c$$
$$= \int \frac{z_c^{L/2}}{(2\pi\sigma_c^2)^{(W_c+L)/2}} \exp\left(-z_c \frac{\hat{\boldsymbol{\varepsilon}}'_c\ddot{\mathbf{W}}_c\hat{\boldsymbol{\varepsilon}}_c + \hat{\mathbf{u}}'_c\hat{\mathbf{u}}_c + (\mathbf{u}_c - \hat{\mathbf{u}}_c)'\mathbf{M}_{c(\mathbf{u})}(\mathbf{u}_c - \hat{\mathbf{u}}_c)}{2\sigma_c^2}\right) d\mathbf{u}_c, \quad (A15)$$
$$= \frac{z_c^{L/2}\exp\left(-\frac{\hat{\boldsymbol{\varepsilon}}'_c\ddot{\mathbf{W}}_c\hat{\boldsymbol{\varepsilon}}_c + \hat{\mathbf{u}}'_c\hat{\mathbf{u}}_c}{2\sigma_c^2 z_c^{-1}}\right)}{z_c^{L/2}(2\pi\sigma_c^2)^{W_c/2}|\mathbf{M}_{c(\mathbf{u})}|^{1/2}} \int \frac{z_c^{L/2}|\mathbf{M}_{c(\mathbf{u})}|^{1/2}}{(2\pi\sigma_c^2)^{L/2}} \exp\left(-\frac{(\mathbf{u}_c - \hat{\mathbf{u}}_c)'\mathbf{M}_{c(\mathbf{u})}(\mathbf{u}_c - \hat{\mathbf{u}}_c)}{2\sigma_c^2 z_c^{-1}}\right) d\mathbf{u}_c.$$

The integral in the third line equals one because it integrates a Gaussian PDF ($\mathbf{u}_c \sim N(\hat{\mathbf{u}}_c, z_c^{-1}\sigma_c^2 \mathbf{M}_{c(\mathbf{u})}^{-1})$). Eventually, we have the following closed-form expression for the $c$-th sub-model likelihood:

$$L(\sigma_c^2, \boldsymbol{\theta}_c, \mathbf{b}_c) = \frac{1}{(2\pi\sigma_c^2)^{W_c/2}|\mathbf{M}_{c(\mathbf{u})}|^{1/2}} \exp\left(-\frac{\hat{\boldsymbol{\varepsilon}}'_c\ddot{\mathbf{W}}_c\hat{\boldsymbol{\varepsilon}}_c + \hat{\mathbf{u}}'_c\hat{\mathbf{u}}_c}{2\sigma_c^2 z_c^{-1}}\right), \quad (A16)$$

The likelihood of the aggregated model can be expanded as $L(\sigma_{1:C}^2, \boldsymbol{\theta}_{1:C}) = \int f(\mathbf{y}, \mathbf{u}_{1:C}|\mathbf{b}_{1:C}, \sigma_{1:C}^2, \boldsymbol{\theta}_{1:C})\, d\mathbf{u}_{1:C} = \int \prod_{c=1}^C f(\mathbf{y}_c, \mathbf{u}_c|\mathbf{b}_c, \sigma_c^2, \boldsymbol{\theta}_c)\, d\mathbf{u}_{1:C} = \prod_{c=1}^C f(\mathbf{y}_c|\mathbf{b}_c, \sigma_c^2, \boldsymbol{\theta}_c)$ . Using this, we have the following log-likelihood:

$$\log L(\sigma_{1:C}^2, \boldsymbol{\theta}_{1:C}, \mathbf{b}_{1:C}) = \sum_{c=1}^C \left(-\frac{W_c}{2}\log(2\pi\sigma_c^2) - \frac{1}{2}\log|\mathbf{M}_{c(\mathbf{u})}| - \frac{\hat{\boldsymbol{\varepsilon}}'_c\ddot{\mathbf{W}}_c\hat{\boldsymbol{\varepsilon}}_c + \hat{\mathbf{u}}'_c\hat{\mathbf{u}}_c}{2\sigma_c^2 z_c^{-1}}\right). \quad (A17)$$

Since $\frac{d\log L(\sigma_c^2, \boldsymbol{\theta}_c)}{d\sigma_c^2} = \frac{d}{d\sigma_c^2}\left(-\frac{W_c}{2}\log(2\pi\sigma_c^2) - \frac{\hat{\boldsymbol{\varepsilon}}'_c\ddot{\mathbf{W}}_c\hat{\boldsymbol{\varepsilon}}_c + \hat{\mathbf{u}}'_c\hat{\mathbf{u}}_c}{2\sigma_c^2 z_c^{-1}}\right) = -\frac{W_c}{2\sigma_c^2} + \frac{(\hat{\boldsymbol{\varepsilon}}'_c\ddot{\mathbf{W}}_c\hat{\boldsymbol{\varepsilon}}_c + \hat{\mathbf{u}}'_c\hat{\mathbf{u}}_c)}{2\sigma_c^4 z_c^{-1}}$, the estimate of $\sigma_c^2$ maximizing Eq. (A17) is $\hat{\sigma}_c^2 = \frac{\hat{\boldsymbol{\varepsilon}}'_c\ddot{\mathbf{W}}_c\hat{\boldsymbol{\varepsilon}}_c + \hat{\mathbf{u}}'_c\hat{\mathbf{u}}_c}{N_c}$. By substituting it, the following marginal log-likelihood (note: $\frac{\hat{\boldsymbol{\varepsilon}}'_c\ddot{\mathbf{W}}_c\hat{\boldsymbol{\varepsilon}}_c + \hat{\mathbf{u}}'_c\hat{\mathbf{u}}_c}{2\hat{\sigma}_c^2 z_c^{-1}} = \frac{N_c}{2(W_c/N_c)^{-1}} = \frac{W_c}{2}$) is obtained:

$$\log L(\boldsymbol{\theta}_{1:C}, \mathbf{b}_{1:C}) = \sum_{c=1}^C \left(-\frac{W_c}{2}\log\left(2\pi \frac{\hat{\boldsymbol{\varepsilon}}'_c\ddot{\mathbf{W}}_c\hat{\boldsymbol{\varepsilon}}_c + \hat{\mathbf{u}}'_c\hat{\mathbf{u}}_c}{N_c}\right) - \frac{1}{2}\log|\mathbf{M}_{c(\mathbf{u})}| - \frac{W_c}{2}\right). \quad (A18)$$

After replacing $\mathbf{b}_c$ with $\hat{\mathbf{b}}_c$, $\boldsymbol{\theta}_{1:C}$ can be estimated by maximizing Eq. (A18).

---

$\hat{\mathbf{u}}_c)'\frac{\partial^2 \log f(\mathbf{y}_c,\mathbf{u}_c|\mathbf{b}_c,\sigma_c^2,\boldsymbol{\theta}_c)}{\partial \mathbf{u}_c \partial \mathbf{u}'_c}(\mathbf{u}_c - \hat{\mathbf{u}}_c)$ where $\frac{\partial^2 \log f(\mathbf{y}_c,\mathbf{u}_c|\mathbf{b}_c,\sigma_c^2,\boldsymbol{\theta}_c)}{\partial \mathbf{u}_c \partial \mathbf{u}'_c} = \mathbf{M}_{c(\mathbf{u})}$ and higher order terms are zero since $\mathbf{y}_c$ and $\mathbf{u}_c$ are Gaussians (see Wood, 2017). By organizing the equality, we have $\boldsymbol{\varepsilon}'_c\ddot{\mathbf{W}}_c\boldsymbol{\varepsilon}_c + \mathbf{u}'_c\mathbf{u}_c = \hat{\boldsymbol{\varepsilon}}'_c\ddot{\mathbf{W}}_c\hat{\boldsymbol{\varepsilon}}_c + \hat{\mathbf{u}}'_c\hat{\mathbf{u}}_c + (\mathbf{u}_c - \hat{\mathbf{u}}_c)'\mathbf{M}_{c(\mathbf{u})}(\mathbf{u}_c - \hat{\mathbf{u}}_c)$.

**Appendix 3: Derivation of the marginal log-likelihood:** $l(\boldsymbol{\theta}_{1:C}) = \log L(\boldsymbol{\theta}_{1:C})$

Following Bates (2010), in the same way as Eq. (A14), the following equality holds:

$$\boldsymbol{\varepsilon}'_c \ddot{\mathbf{W}}_c \boldsymbol{\varepsilon}_c + \mathbf{u}'_c \mathbf{u}_c = \hat{\boldsymbol{\varepsilon}}'_c \ddot{\mathbf{W}}_c \hat{\boldsymbol{\varepsilon}}_c + \hat{\mathbf{u}}'_c \hat{\mathbf{u}}_c + (\mathbf{q}_c - \hat{\mathbf{q}}_c)' \mathbf{M}_c (\mathbf{q}_c - \hat{\mathbf{q}}_c), \tag{A19}$$

where $\mathbf{q}_c = [\mathbf{b}'_c, \mathbf{u}'_c]'$ and $\hat{\boldsymbol{\varepsilon}}_c = \mathbf{y}_c - \mathbf{X}_c \hat{\mathbf{b}}_c - \mathbf{E}_c \mathbf{V}_c(\hat{\boldsymbol{\theta}}_c) \hat{\mathbf{u}}_c = \mathbf{y}_c - [\mathbf{X}_c, \mathbf{E}_c \mathbf{V}_c(\hat{\boldsymbol{\theta}}_c)] \hat{\mathbf{q}}_c$ in this section. $\mathbf{M}_c = \frac{\partial^2 \log f(\mathbf{y}_c, \mathbf{u}_c | \mathbf{b}_c, \sigma_c^2, \boldsymbol{\theta}_c)}{\partial \mathbf{q}_c \partial \mathbf{q}'_c}$. By applying the same expansion as Eqs. (A15-A16),

$$\begin{aligned}
L(\sigma_c^2, \boldsymbol{\theta}_c) &= \int f(\mathbf{y}_c, \mathbf{u}_c | \hat{\mathbf{b}}_c, \sigma_c^2, \boldsymbol{\theta}_c) \, \mathbf{q}_c \\
&= \int \frac{z_c^{L/2}}{(2\pi\sigma_c^2)^{(W_c + L)/2}} \exp\left(-z_c \frac{\hat{\boldsymbol{\varepsilon}}'_c \ddot{\mathbf{W}}_c \hat{\boldsymbol{\varepsilon}}_c + \hat{\mathbf{u}}'_c \hat{\mathbf{u}}_c + (\mathbf{q}_c - \hat{\mathbf{q}}_c)' \mathbf{M}_c (\mathbf{q}_c - \hat{\mathbf{q}}_c)}{2\sigma_c^2}\right) d\mathbf{q}_c, \\
&= \frac{z_c^{L/2} \exp\left(-\frac{\hat{\boldsymbol{\varepsilon}}'_c \ddot{\mathbf{W}}_c \hat{\boldsymbol{\varepsilon}}_c + \hat{\mathbf{u}}'_c \hat{\mathbf{u}}_c}{2\sigma_c^2 z_c^{-1}}\right)}{z_c^{(K+L)/2} (2\pi\sigma_c^2)^{(W_c - K)/2} |\mathbf{M}_c|^{1/2}} \times \\
&\quad \int \frac{z_c^{(K+L)/2} |\mathbf{M}_c|^{1/2}}{(2\pi\sigma_c^2)^{(K+L)/2}} \exp\left(-\frac{(\mathbf{q}_c - \hat{\mathbf{q}}_c)' \mathbf{M}_c (\mathbf{q}_c - \hat{\mathbf{q}}_c)}{2\sigma_c^2 z_c^{-1}}\right) d\mathbf{v}_c \\
&= \frac{1}{z_c^{K/2} (2\pi\sigma_c^2)^{(W_c - K)/2} |\mathbf{M}_c|^{1/2}} \exp\left(-\frac{\hat{\boldsymbol{\varepsilon}}'_c \ddot{\mathbf{W}}_c \hat{\boldsymbol{\varepsilon}}_c + \hat{\mathbf{u}}'_c \hat{\mathbf{u}}_c}{2\sigma_c^2 z_c^{-1}}\right)
\end{aligned} \tag{A20}$$

The marginal likelihood of the aggregated model is expanded as $L(\sigma_{1:C}^2, \boldsymbol{\theta}_{1:C}) = \int f(\mathbf{y}, \mathbf{u}_{1:C} | \mathbf{b}_{1:C}, \sigma_{1:C}^2, \boldsymbol{\theta}_{1:C}) d\mathbf{u}_{1:C} = \int \prod_{c=1}^{C} f(\mathbf{y}_c, \mathbf{u}_c | \mathbf{b}_c, \sigma_c^2, \boldsymbol{\theta}_c) d\mathbf{u}_{1:C} = \prod_{c=1}^{C} f(\mathbf{y}_c | \mathbf{b}_c, \sigma_c^2, \boldsymbol{\theta}_c)$.
Using this, we have the following log-likelihood we will maximize in the ML estimation as follows:

$$\log L(\sigma_{1:C}^2, \boldsymbol{\theta}_{1:C}) = \sum_{c=1}^{C} \left(-\frac{K}{2} z_c - \frac{W_c - K}{2} \log(2\pi\sigma_c^2) - \frac{1}{2} \log|\mathbf{M}_c| - \frac{\hat{\boldsymbol{\varepsilon}}'_c \ddot{\mathbf{W}}_c \hat{\boldsymbol{\varepsilon}}_c + \hat{\mathbf{u}}'_c \hat{\mathbf{u}}_c}{2\sigma_c^2 z_c^{-1}}\right) \tag{A21}$$

Since $\frac{d \log L(\sigma_c^2, \boldsymbol{\theta}_c)}{d\sigma_c^2} = \frac{d}{d\sigma_c^2}\left(-\frac{W_c - K}{2} \log(2\pi\sigma_c^2) - \frac{\hat{\boldsymbol{\varepsilon}}'_c \ddot{\mathbf{W}}_c \hat{\boldsymbol{\varepsilon}}_c + \hat{\mathbf{u}}'_c \hat{\mathbf{u}}_c}{2\sigma_c^2 z_c^{-1}}\right) = -\frac{W_c - K}{2\sigma_c^2} + \frac{(\hat{\boldsymbol{\varepsilon}}'_c \ddot{\mathbf{W}}_c \hat{\boldsymbol{\varepsilon}}_c + \hat{\mathbf{u}}'_c \hat{\mathbf{u}}_c)}{2\sigma_c^4 z_c^{-1}}$, the

estimate of $\sigma_c^2$ maximizing Eq. (A21) is $\hat{\sigma}_c^2 = \frac{\hat{\boldsymbol{\varepsilon}}'_c \ddot{\mathbf{W}}_c \hat{\boldsymbol{\varepsilon}}_c + \hat{\mathbf{u}}'_c \hat{\mathbf{u}}_c}{N_c - K}$. By substituting it, we have the following marginal log-likelihood (note: $-\frac{K}{2} z_c - \frac{\hat{\boldsymbol{\varepsilon}}'_c \ddot{\mathbf{W}}_c \hat{\boldsymbol{\varepsilon}}_c + \hat{\mathbf{u}}'_c \hat{\mathbf{u}}_c}{2\hat{\sigma}_c^2 z_c^{-1}} = -\frac{KW_c + (N_c - K)W_c}{2N_c} = -\frac{W_c}{2}$):

$$\log L(\boldsymbol{\theta}_{1:C}) = \sum_{c=1}^{C} \left(-\frac{W_c - K}{2} \log\left(2\pi \frac{\hat{\boldsymbol{\varepsilon}}'_c \ddot{\mathbf{W}}_c \hat{\boldsymbol{\varepsilon}}_c + \hat{\mathbf{u}}'_c \hat{\mathbf{u}}_c}{N_c - K}\right) - \frac{1}{2} \log|\mathbf{M}_c| - \frac{W_c}{2}\right). \tag{A22}$$

It equals the marginal log-likelihood Eq. (22) we will maximize.

Appendix 4: Preliminary analysis on the number of sub-samples/clusters

This section evaluates SVC estimation accuracy of ESF-MA$_{GL}$ while varying the average sample size per cluster/sub-sample as $N_c \in \{100, 200, 400, 600, 800\}$. $N = 70^2$ and $r = 1$ are assumed. Data generating process and other settings are the same as Section 5.1.

RMSEs and computation time are summarized in Figures A1 and A2 respectively, together with those for the original (RE)ESF. Remember that the standard deviation of the strong SVC is twice the standard deviation of the weak SVC. Compared to the basic ESF, ESF-MA$_{GL}$ has similar or larger RMSE values for the weak SVCs when $N_c \leq 400$ while smaller RMSEs when $600 \leq N_c$. The computation time is acceptable even if $N_c = 600$. Because of these reasons, we assumed $N_c = 600$ in this study.

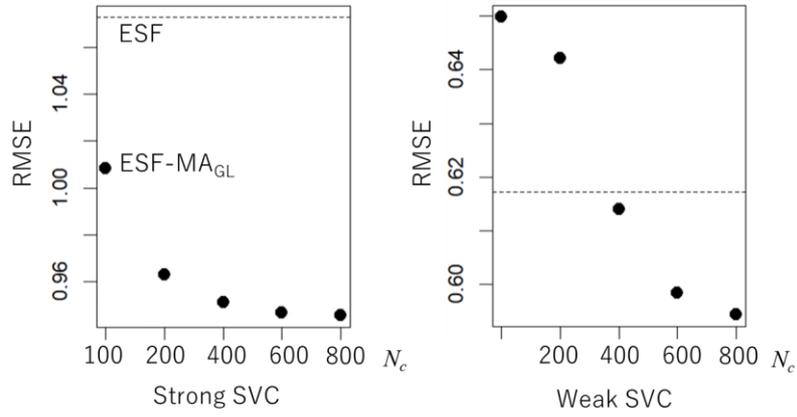

Figure A1: SVC estimation accuracy while varying the sample size $N_c$ per cluster.

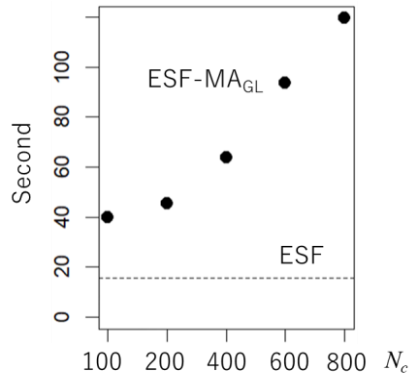

Figure A2: Computation time while varying the sample size $N_c$ per cluster. Both the computations are unparalleled.